%% file: neurips_2026.tex
\documentclass{article}

\PassOptionsToPackage{numbers, compress}{natbib}

\usepackage[preprint]{neurips_2026}
\addtocontents{toc}{\protect\setcounter{tocdepth}{-1}}

\usepackage{amsmath}
\usepackage{algorithm}
\usepackage{algpseudocode}
\usepackage{booktabs}
\usepackage[table]{xcolor}
\definecolor{paramgray}{RGB}{235,235,235}
\definecolor{highlightrow}{RGB}{230,239,238}
 \usepackage{amssymb}
\usepackage{multirow}
\usepackage{graphicx}
\usepackage{subcaption}

\usepackage[utf8]{inputenc} % allow utf-8 input
\usepackage[T1]{fontenc}    % use 8-bit T1 fonts
\usepackage{hyperref}       % hyperlinks
\usepackage{url}            % simple URL typesetting
\usepackage{booktabs}       % professional-quality tables
\usepackage{amsfonts}       % blackboard math symbols
\usepackage{nicefrac}       % compact symbols for 1/2, etc.
\usepackage{microtype}      % microtypography
\usepackage{pifont}         % \ding{51} (check) and \ding{55} (cross)
\usepackage{xspace}         % \xspace for command names that need trailing space
\usepackage{xcolor}         % colors
\usepackage{tikz}           % vector graphics for case study
\usetikzlibrary{positioning,arrows.meta,fit,backgrounds}
\usepackage[normalem]{ulem} % \sout for strikethrough in worked example
\usepackage[most]{tcolorbox}% boxed text/code blocks for case study

\usepackage{booktabs}
\usepackage[table]{xcolor}
\definecolor{paramgray}{RGB}{235,235,235}
\definecolor{highlightrow}{RGB}{230,239,238}

\newcommand{\methodname}{P2T\xspace}

\title{From Patches to Trajectories: Privileged Process Supervision for Software-Engineering Agents}

\author{%
  Murong Ma\textsuperscript{1}\thanks{Work done during internship at Microsoft Research.} \quad
  Tianyu Chen\textsuperscript{2}\thanks{Corresponding authors: \texttt{chentianyu@microsoft.com}, \texttt{lin\_yun@sjtu.edu.cn}, \texttt{yegong@microsoft.com}.} \quad
  Yun Lin\textsuperscript{3}\footnotemark[2] \quad
  Shuai Lu\textsuperscript{2} \quad
  Qinglin Zhu\textsuperscript{4} \\
  \textbf{Yeyun Gong\textsuperscript{2}}\footnotemark[2] \quad
  \textbf{Zhiyong Huang\textsuperscript{1}} \quad
  \textbf{Peng Cheng\textsuperscript{2}} \quad
  \textbf{Yan Lu\textsuperscript{2}} \quad
  \textbf{Jin Song Dong\textsuperscript{1}} \\[2pt]
  \textsuperscript{1}National University of Singapore \quad
  \textsuperscript{2}Microsoft Research Asia \\
  \textsuperscript{3}Shanghai Jiao Tong University \quad
  \textsuperscript{4}King's College London
}

\begin{document}

\maketitle

\input{sections/abstract}
\input{sections/introduction}

\input{sections/related_work}

\input{sections/problem}
\input{sections/method}

\input{sections/experiments}

\input{sections/conclusion}

\bibliography{example_paper}
\bibliographystyle{icml2026}

\input{sections/appendix}
\end{document}

%% file: sections/abstract.tex
\begin{abstract}
Supervised fine-tuning (SFT) on long teacher trajectories is the dominant method for instilling investigation and reasoning capabilities into open software-engineering (SWE) agents. Under SFT, every retained response is an imitation target, so the student inherits not only the trajectory's outcome but also any flaw in its intermediate steps, including ungrounded leaps and redundant loops. High-quality training data must therefore be jointly \emph{effective} (each step is grounded and narrows the agent's epistemic gap to the correct fix) and \emph{efficient} (each step is information-bearing rather than redundant or looping). Existing recipes filter or relabel teacher rollouts using only a binary terminal verifier, which does not directly target these axes and provides no supervision on instances where the teacher fails.

Every real issue ships with a developer-authored reference patch $p^\star$ that implicitly testifies to the file paths, runtime behaviors, and conventions a fix presupposes, but the standard pipeline discards it. We propose \textbf{\methodname} (Patches-to-Trajectories), which uses $p^\star$ as \emph{privileged information} during curation, and frames trajectory construction as a bi-objective program over per-step effectiveness and trajectory length. A \emph{reverse} phase distills $p^\star$ into a latent \emph{process graph} $G^\star$ of contextual facts and solution milestones, encoding dense intermediate anchors in constructive ordering. A \emph{forward} phase curate trajectories from blinded teacher continuations, scoring per-step progress against $G^\star$ under a leakage-blocking groundedness check and committing the shortest segments that retain effectiveness.

 Using only $1.8$k curated SWE-Gym instances, \methodname improves both axes simultaneously over outcome-filtered SFT and its tool-error-masking variant: on SWE-bench Verified, it lifts Pass@1 by up to $+10.8$ points while cutting per-instance inference cost by $\sim\!15\%$, with consistent gains on SWE-bench Lite and across two teachers. A size-matched ablation and qualitative analysis further isolate per-trajectory quality from data scale.
\end{abstract}

%% file: sections/introduction.tex
\section{Introduction}

Autonomous software-engineering agents built on large language models (LLMs) are now routinely competitive on real GitHub issue-resolution benchmarks~\citep{jimenez2024swebench, deng2025swepro, zhang2025swelive, zan2025swemulti}, navigating repositories, localizing faults, editing code, and validating fixes~\citep{yang2024sweagent,wang2024openhands}. A capable agent must do more than emit a final patch that happens to pass: it must learn to \emph{investigate}, \emph{reason}, and \emph{validate}, the per-step competencies that make terminal success reproducible rather than incidental. The dominant route to instilling these competencies in open base models is supervised fine-tuning (SFT) on long trajectories from strong teacher models~\citep{ouyang2022training,pan2024swegym,jain2025r2egym,zeng2025skyworkswe}, which provides dense process supervision across turns of ReAct-style interactions~\citep{yao2023react}. Under SFT---a behavior-cloning objective for sequential decision problems~\citep{ross2011reduction}---every retained response is an imitation target, so the student inherits not only the trajectory's outcome but also any flaw in its intermediate steps. Each trajectory therefore needs two complementary properties. \textbf{Effectiveness:} each step narrows the agent's epistemic gap to the correct fix by uncovering a fact the fix presupposes, grounded in the visible prefix with no unsupported leaps or premature conclusions. \textbf{Efficiency:} each step is information-bearing, advancing the trajectory rather than re-deriving established facts, looping on uninformative actions, or padding with redundant exploration. The two are in direct tension: cautious exploration lengthens trajectories, while aggressive shortening invites unsupported shortcuts. Constructing trajectories on the right side of this tradeoff is the central data problem for SFT of SWE agents.

Existing recipes do not directly target these axes. The standard pipeline samples teacher rollouts and retains only those whose final patch passes the issue's tests~\citep{pan2024swegym}; variants scale the instance pool by procedurally synthesizing executable issues~\citep{yang2025swesmith,jain2025r2egym,zeng2025skyworkswe}. All inherit the same binary terminal verifier: an outcome-supervision signal that supplies feedback only on the final result rather than on intermediate reasoning steps~\citep{lightman2024lets}. It is therefore structurally indifferent to either axis. On the SWE-Gym training pool, retained trajectories often exhaust the $100$-iteration budget without reaching a normal finish, accidentally tripping the test suite ($7.6\%$ under the Qwen3-Coder-480B teacher, $9.3\%$ under GLM-5-FP8); $6.8\%$/$8.6\%$ of their file-viewing actions revisit content already viewed earlier in the trajectory; and $70.2\%$/$64.7\%$ of instances contribute \emph{no} supervision because the teacher never produces a passing patch. Independent audits further show that a non-trivial share of ``passing'' patches reflect weak tests or solution leakage rather than correctness~\citep{aleithan2024swebenchplus}; more broadly, test-suite-based program repair has long been known to admit plausible but incorrect or overfitted patches~\citep{qi2015analysis,smith2015cure}. The verifier is therefore not even tight on its own axis. The remaining question is therefore not how to acquire \emph{more} tasks, but how to extract \emph{better} per-trajectory supervision from the real ones.

The signal that addresses both axes is already available, but unused: the developer-authored \emph{reference patch} $p^\star$ associated with each issue--pull-request instance in real-issue SWE benchmarks~\citep{jimenez2024swebench}, which enters the standard pipeline only as a discarded ground truth. As process supervision, $p^\star$ is uniquely well-positioned, since each line of it implicitly testifies to the file paths, runtime behaviors, and conventions a solver would have had to uncover before the edit becomes derivable. We therefore propose to use $p^\star$ as \emph{privileged information}~\citep{vapnik2009lupi} during curation: a quantity the data-construction procedure may consult to score and shape trajectories, but that the student never sees. With $p^\star$ in scope, the curator can score per-step progress against the prerequisites a fix presupposes (effectiveness), keep only information-bearing steps (efficiency), and recover supervision precisely on the hard instances where ordinary teacher rollouts fail.

The challenge is that conditioning trajectory generation directly on $p^\star$ leaks the answer: any prefix built with $p^\star$ in scope risks splicing in edits, claims, or file references no honest investigation could yet support, and a student that imitates such a trace internalizes the same unjustified leaps. We therefore propose \textbf{\methodname} (\emph{Patches-to-Trajectories}), which frames trajectory curation as a \emph{bi-objective program} over per-step effectiveness and trajectory length, and mediates $p^\star$ through a latent \emph{process graph} $G^\star$ distilled from it, so the curator can shape trajectories along both axes without ever exposing $p^\star$ to the student.
Empirically, \methodname improves both axes simultaneously over outcome-filtered SFT and its tool-error-masking variant: on SWE-bench Verified, it lifts Pass@1 by up to $+10.8$ points while cutting per-instance inference cost by $\sim\!15\%$, and a size-matched control already beats both baselines on both axes, isolating per-trajectory quality from data scale. % Full experimental details are deferred to Sec.~\ref{sec:experiments}.
Our contributions are as follows:

% \paragraph{Contributions.}
\begin{itemize}\setlength{\itemsep}{1pt}
    \item We frame SWE-agent SFT data construction as a \emph{bi-objective} program over per-step effectiveness and trajectory length, and show that outcome-filtered rejection sampling provides no per-step or length signal.
    \item We propose \methodname, a curation framework that uses $p^\star$ as \emph{privileged information}: a reverse phase distills $p^\star$ into a process graph $G^\star$ of contextual facts and solution milestones, and a forward phase realizes trajectories that are short, grounded, and steered by $G^\star$.
    \item Using only $1.8$k curated SWE-Gym trajectories, \methodname improves Pass@1 by up to $+10.8$ points while cutting per-instance inference cost by $\sim\!15\%$ over outcome-filtered SFT, on SWE-bench Verified and Lite across two students and two teachers.
\end{itemize}

%% file: sections/related_work.tex
\section{Related Work}
\label{sec:related_work}

\textbf{SWE agents and benchmarks.}
SWE-bench~\citep{jimenez2024swebench} catalyzed a line of inference-time systems for repository-level issue resolution: ReAct-style tool use~\citep{yao2023react}, SWE-agent's agent--computer interface~\citep{yang2024sweagent}, the OpenHands platform~\citep{wang2025openhands}, structure-aware retrieval in AutoCodeRover~\citep{zhang2024autocoder}, and the simpler localize--repair--validate pipeline of Agentless~\citep{xia2024agentless}. Audits show that terminal pass/fail can overstate correctness when tests are weak or issues leak the solution~\citep{aleithan2024swebenchplus, wang2025solved}. Our work is orthogonal to these scaffolds: we improve the per-step \emph{quality} of the SFT trajectories on which such agents are trained.

\textbf{Trajectory data for open SWE agents.}
Existing recipes scale executable tasks and retain trajectories that pass a terminal verifier: SWE-Gym~\citep{pan2024swegym} on real Python issues, R2E-Gym~\citep{jain2025r2egym} with procedural construction and hybrid verifiers, SWE-smith~\citep{yang2025swesmith} via test-breaking synthesis, and Skywork-SWE~\citep{zeng2025skyworkswe} on large-scale curation and trajectory scaling. All retain whole successful rollouts, inheriting their detours, redundant observations, and unsupported inferences. \methodname instead treats $p^\star$ as privileged curation information, distilled into $G^\star$ to expose only the prerequisites a fix presupposes, while never showing $p^\star$ to the student.

% \textbf{Process supervision.}
% Replacing final-only labels with step-level feedback improves reasoning models on math, via human annotation~\citep{lightman2024lets} or automatic Monte Carlo estimation~\citep{wang2024mathshepherd}; analogous step rewards have been applied to general agents~\citep{xiong2024watchevery} and web navigation~\citep{chae2025webshepherd}. These methods presuppose annotatable steps or a queryable process reward. SWE agents have neither: tests are sparse and noisy, while $p^\star$ is dense but leakage-prone. \methodname operationalizes $p^\star$ through $G^\star$ to obtain per-step supervision without exposing the answer.

\textbf{Trajectory and context reduction for LLM agents.}
A complementary line attacks the inference-time cost of long agent histories: AgentDiet prunes useless, redundant, or expired entries from coding-agent trajectories at run time~\citep{xiao2026agentdiet}, while ACON learns to compress observations and interaction histories for long-horizon agents~\citep{kang2025acon}. These methods leave the underlying policy fixed and shorten what it consumes; \methodname instead shortens what it \emph{produces} at training time, so the resulting student is intrinsically efficient and remains compatible with such inference-time compressors.

%% file: sections/problem.tex
\section{Problem Setting}
\label{sec:problem}

We study the construction of process-supervision data for supervised fine-tuning (SFT) of autonomous software-engineering agents. Under SFT, every response in a training trajectory becomes an imitation target, so a trajectory's training value is bounded by its weakest step: a passing terminal patch does not redeem a prefix that contains hallucinated reasoning, redundant exploration, non-progressing action loops, or uninformative tool calls. Data construction must therefore control two complementary properties of each trajectory: \emph{process effectiveness}---every step makes prefix-grounded progress toward the reference fix, and \emph{process efficiency}---the trajectory is short, to limit student inference cost and reduce the surface area of imitation noise. The two are in direct tension: cautious, well-grounded exploration lengthens trajectories, while aggressive shortening invites unsupported leaps. We frame trajectory curation as a \emph{bi-objective optimization} problem over these two criteria.

\subsection{Trajectories and the SFT Objective}
\label{sec:problem_general}

A task instance is a tuple
\(
\mathcal{I}_i = (d_i,\, R_i,\, E_i,\, \mathcal{T}_i,\, p_i^\star),
\)
where \(d_i\) is the issue description, \(R_i\) the repository at the pre-fix commit, \(E_i\) a sandboxed execution environment exposing a fixed tool set (file viewing, shell execution, code editing), \(\mathcal{T}_i\) the issue's test suite, and \(p_i^\star\) the reference fix patch. We split the components of \(\mathcal{I}_i\) into the \emph{non-oracle bundle}
\(
x_i := (d_i, R_i, E_i)
\)
that any solver may consume, and the \emph{oracle bundle} \((p_i^\star, \mathcal{T}_i)\), which the data-construction procedure may use to evaluate trajectories.

The agent interacts with \(R_i\) through \(E_i\). At turn \(t\) it observes the visible prefix
\(
h_t = (d_i, y_1, o_1, \ldots, y_{t-1}, o_{t-1})
\)
and emits a ReAct-style response \(y_t = (c_t, a_t)\), comprising reasoning \(c_t\) and an action \(a_t\); executing \(a_t\) in \(E_i\) returns an observation \(o_t\). A trajectory is the resulting sequence
\(
\tau_i = (d_i,\, y_1, o_1,\, \ldots,\, y_T, o_T),
\)
and a constructed collection induces the SFT dataset \(\mathcal{D} = \{(h_t, y_t) : (h_t, y_t) \in \tau_i\}\). The student \(\pi_\theta\) is trained by behavioral cloning,
\[
\mathcal{L}_{\mathrm{SFT}}(\theta) \;=\; -\,\mathbb{E}_{(h_t, y_t) \sim \mathcal{D}}\bigl[\log \pi_\theta(y_t \mid h_t)\bigr].
\]
Because every retained response becomes a training target, data construction must control not only whether the final patch is correct, but whether the intermediate process is itself worth imitating.

\subsection{Outcome-Filtered Trajectory Collection}
\label{sec:problem_rejection}

The dominant paradigm uses \(\mathcal{T}_i\) purely as a terminal verifier and discards \(p_i^\star\). Given a teacher policy \(\pi_T\), one samples \(K\) trajectories \(\tau_i^{(k)} \sim \pi_T(\cdot \mid x_i)\) per instance and retains those whose induced patch \(\hat p_i^{(k)} = \mathrm{patch}(\tau_i^{(k)})\) passes the test suite:
\[
\mathcal{D}_{\mathrm{OF}}
\;=\;
\bigl\{\,(h_t, y_t) \in \tau_i^{(k)} \;:\; \mathcal{T}_i\bigl(\hat p_i^{(k)}\bigr) = 1\,\bigr\}.
\]
This procedure treats terminal success as the only supervision signal. As a result, retained data might include low-quality trajectories, i.e. failing to distinguish a concise, evidence-driven solution from one that succeeds after redundant search, unsupported claims, or accidental edits.

% Start from here (by tianyu)
% This procedure treats terminal success as the only supervision signal. As a result, retained data may be \emph{Pareto-dominated} by trajectories that produce the same fix through shorter or better-grounded processes: 
% \tianyu{it seems to explain why this task is Pareto-dominated while the rest of this paragraph is the limitation of existing work. I suggest not to mention ``Pareto optimization'' here, just explaining the limitation of existing work. Murong can refer to the following paragraph, we can move the definition of ``Pareto optimization'' either before (in section 3.0) or later (in section 3.3); we can also remove this definition, and using only Bi-objective trajectory optimization.}
% outcome filtering does not distinguish a concise, evidence-driven solution from one that succeeds after redundant search, unsupported claims, or accidental edits. The reference patch \(p_i^\star\), which encodes precisely how a competent developer resolves the issue, is also left unused as a source of process supervision. Both pathologies follow directly from collapsing trajectory quality onto a single binary outcome.

% \tianyu{This procedure treats terminal success as the only supervision signal. As a result, retained data might include low-quality trajectories, i.e. failing to distinguish a concise, evidence-driven solution from one that succeeds after redundant search, unsupported claims, or accidental edits.}
% End from here (by tianyu)

\subsection{Patch-Oracled Bi-objective Trajectory Construction}
\label{sec:problem_ours}
To avoid such low-quality trajectory, in this paper, we propose to utilize the reference patch \(p_i^\star\), which encodes precisely how a competent developer resolves the issue as a source of process supervision.
Specifically, we make \(p_i^\star\) a process oracle: a trajectory is judged by whether its steps uncover the evidence---file and symbol localizations, runtime behavior, and implementation choices---needed to derive a fix equivalent to \(p_i^\star\). This admits any trajectory that establishes the right intermediate evidence, and rejects trajectories whose patch happens to pass.

\textbf{Bi-objective trajectory target.}
We score each trajectory along two axes. \emph{Process effectiveness} \(\mathrm{Eff}_i(\tau) \in [0, 1]\) rewards steps that uncover fix-relevant evidence without leaping ahead of what the prefix supports; we keep it abstract here and instantiate it in Sec.~\ref{sec:method} via a \emph{process graph} \(G_i^\star\) distilled from \(p_i^\star\). \emph{Process efficiency} is the trajectory length in generated response tokens, \(\mathrm{Len}_i(\tau)\). 
When given a set of trajectories \(\mathcal{T}(\mathcal{I}_i)\) on task \(\mathcal{I}_i\), we pick our target one by the \emph{shortest-above-floor} rule: among trajectories whose effectiveness clears a calibrated floor \(\eta_i\), take the shortest (a standard \(\varepsilon\)-constraint scalarization of bi-objective programs~\citep{miettinen1999nonlinear}); the chosen trajectory is admitted into the SFT dataset only if its final patch passes the test suite \(\mathcal{T}_i\):
\[
\tau_i^\star
\;=\;
\arg\min_{\tau \in \mathcal{T}(\mathcal{I}_i)} \mathrm{Len}_i(\tau)
\quad\text{s.t.}\quad
\mathrm{Eff}_i(\tau) \ge \eta_i,
\qquad
\mathcal{D}_{\mathrm{ours}} \;=\; \{(h_t, y_t) \in \tau_i^\star\}_{i=1}^N.
\]

\textbf{Open challenges.}
This formulation leaves two questions for Sec.~\ref{sec:method}: (i) how to anchor \emph{process effectiveness} so that it captures fix-relevant progress, is sensitive to ungrounded leaps, and is not itself a leakage channel for \(p_i^\star\); and (ii) how to operationalize the resulting bi-objective program tractably, given that \(\mathcal{T}(\mathcal{I}_i)\) cannot be searched exhaustively. 

%% file: sections/method.tex
\section{Method}
\label{sec:method}

\begin{figure}[t]
    \centering
    \makebox[\linewidth][c]{%
        \includegraphics[width=1.05\linewidth]{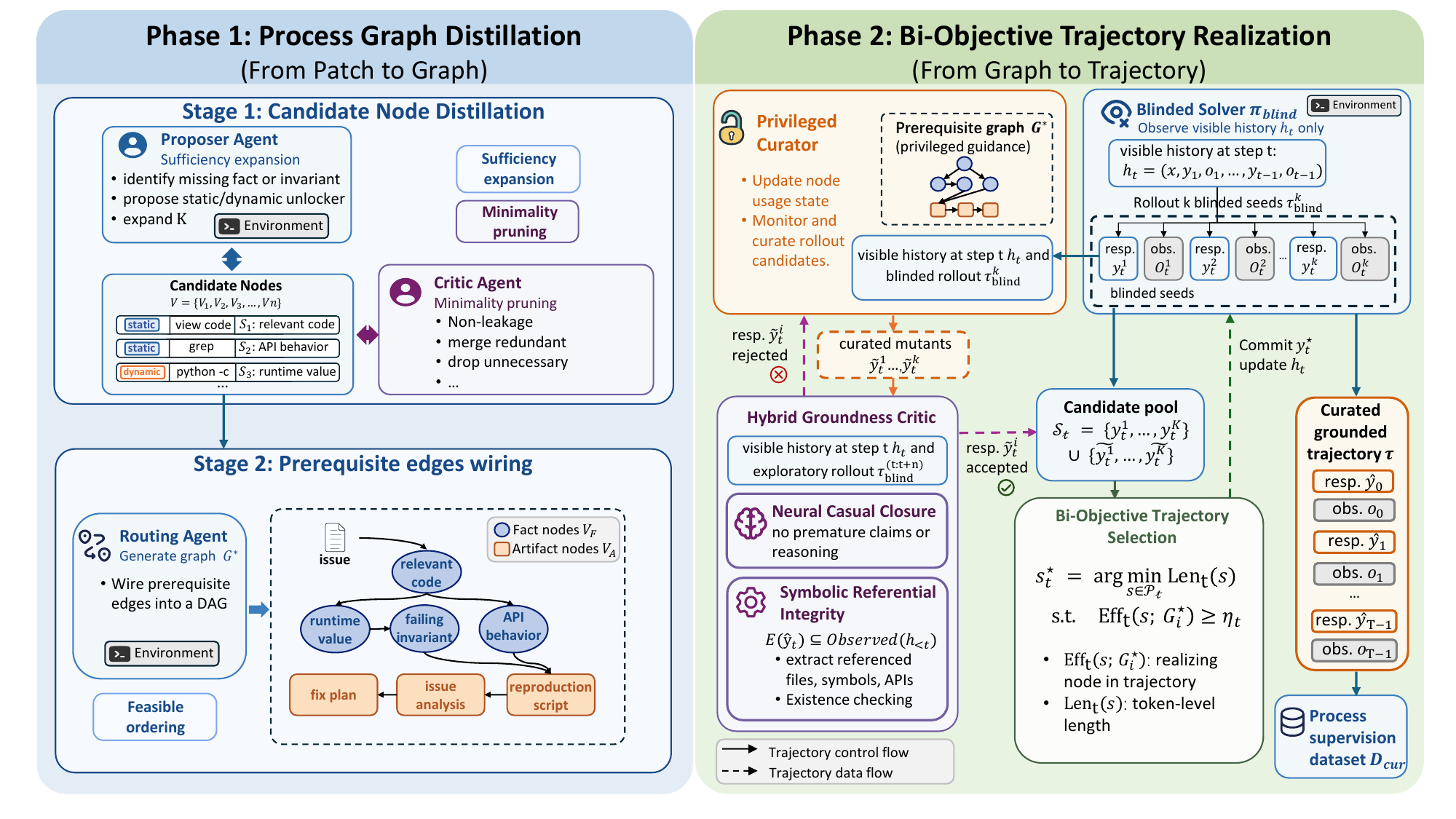}%
    }
    \caption{Overview of \methodname. Phase~1 (Sec.~\ref{sec:reverse}) distills the reference patch \(p_i^\star\) into a minimal latent process graph \(G_i^\star\) of contextual facts and solution milestones---including the eventual edits and validations---that any solver would need to uncover. Phase~2 (Sec.~\ref{sec:forward}) realizes a trajectory by sampling blinded continuations, locally mutating them toward currently available graph nodes, and committing the segment that achieves both effectiveness and efficiency (defined in Sec.~\ref{sec:problem_ours}).}
    \label{fig:method_overview}
    \vspace{-0.2cm}
\end{figure}

We instantiate the bi-objective program of Section~\ref{sec:problem_ours} with a two-phase pipeline that resolves its two open challenges in turn:
\begin{itemize}\setlength{\itemsep}{1pt}
    \item \textbf{Phase 1: Process Graph Distillation (Sec.~\ref{sec:reverse}).} We distill \(p_i^\star\) into a latent \emph{process graph} \(G_i^\star\) whose nodes name the intermediate contextual facts and solution milestones that must be established before the fix becomes derivable, and on top of \(G_i^\star\) we define a \emph{progress} score that measures how much of the graph a trajectory legitimately uncovers---this resolves challenge~(i).
    % \item \textbf{Phase 2: Receding-Horizon Pareto Realization (Sec.~\ref{sec:forward}).} We approximate the trajectory-level bi-objective optimization by \emph{greedy receding-horizon local Pareto selection}: at each window we sample candidate length-\(n\) segments by blind seeding plus optional graph-aware mutation, assemble \(\mathrm{Eff}_t\) by combining the \emph{progress} score with a binary groundedness gate that guards against leakage in the curator's edit, retain the locally non-dominated set under \((\mathrm{Eff}_t, \mathrm{Len}_t)\), commit the shortest segment whose effectiveness clears a threshold, and replan from the new prefix. This resolves challenge~(ii).
    \item \textbf{Phase 2: Receding-Horizon Bi-Objective Trajectory Realization (Sec.~\ref{sec:forward}).} We grow the trajectory one segment at a time via a sliding window. Within each window, we sample a set of candidate segments and apply the same shortest-above-floor rule from Sec.~\ref{sec:problem_ours} locally, then repeat on the extended prefix---this resolves challenge~(ii).
\end{itemize}

Figure~\ref{fig:method_overview} illustrates the pipeline.

\subsection{Phase 1: Process Graph Distillation}
\label{sec:reverse}

The reference patch \(p_i^\star\) defines the \emph{target state} of the repository, but does not itself describe a valid \emph{discovery process}: conditioning trajectory generation directly on \(p_i^\star\) would let edits and unsupported leaps appear in the prefix before the agent has any evidence to justify them. We therefore convert \(p_i^\star\) into a latent process graph
\[
G_i^\star \;=\; (V_i,\, E_i),
\]
in which each node \(v \in V_i\) names an intermediate contextual fact or solution milestone that any solver would need to establish before the fix becomes derivable, and each edge encodes a prerequisite relation. \(G_i^\star\) is the structure on which the per-step effectiveness signal is defined.

\textbf{Node format.}
Each node is represented as
\(
v = (s_v,\, \eta_v,\, u_v),
\)
where \(s_v\) is a natural-language statement, \(\eta_v\) a type tag, and \(u_v\) an explicit \emph{unlocker}: the environment interaction needed to discover \(s_v\). We use two kinds of nodes. \emph{Contextual-fact nodes} record claims about the repository or its runtime behavior whose discovery is a prerequisite for fixing the issue; their unlockers are either \emph{static} (no execution required, e.g., reading a file, inspecting a class hierarchy, or a repository-wide \texttt{grep}) or \emph{dynamic} (require execution, e.g., running a test, evaluating a probe script, or inspecting runtime values). \emph{Solution-milestone nodes} record the intermediate products an agent must construct on the way to the fix---reproduction scripts, root-cause analyses, fix plans, code edits, and validation runs---whose unlockers are the corresponding tool calls (writing a script, drafting an analysis or plan, applying an edit, running the test suite).

\textbf{Three desiderata for \(G_i^\star\).}
We require \(G_i^\star\) to be jointly:
(i) \emph{sufficient}---the issue, repository, and graph nodes together make \(p_i^\star\) plausibly derivable, so that no essential localization, behavioral, or implementation fact is omitted;
(ii) \emph{non-leaking}---each node's unlocker is conceivable from the issue, repository, and the node's predecessors alone, so that proposing the unlocker does not presuppose knowledge of \(p_i^\star\) (e.g.\ an edit node may not appear before its motivating root-cause-analysis and fix-plan nodes);
(iii) \emph{feasibly ordered}---the graph admits a topological order realizable through ordinary environment interaction, with each node discoverable only after its prerequisites are established.

\textbf{Instantiation.}
We construct \(G_i^\star\) by an iterative \emph{proposer--critic} procedure implemented with two specialized LLM agents. Starting from \(V_i^{(0)} = \varnothing\), the proposer adds candidate nodes that close the remaining logical gap to \(p_i^\star\), each annotated with a candidate unlocker; this targets desideratum~(i), \emph{sufficiency}. The critic then prunes any candidate whose declared unlocker is not motivated by the nodes already in \(V_i^{(t)}\)---for instance, an edit node introduced before any root-cause-analysis or fix-plan node---and emits feedback indicating which aspects remain under-determined; this enforces desideratum~(ii), \emph{non-leakage}. The loop terminates when the node set stabilizes. A final organization step then links the surviving nodes into the DAG \(G_i^\star\) by drawing a prerequisite edge from \(u\) to \(v\) whenever \(u\) must be established before \(v\)'s unlocker can apply, enforcing desideratum~(iii), \emph{feasible ordering}. Full prompts are deferred to App.~\ref{app:reverse_prompts}; a worked example on a real SWE-Gym instance is given in App.~\ref{app:worked_example}.

\textbf{Node establishment.}
We say a node \(v\) is \emph{established} by a trajectory prefix \(h_t\), written \(\mathrm{Est}(v, h_t) = 1\), when both: (a) some action in \(h_t\) matches requirement specified by \(u_v\) (e.g., a repository-wide \texttt{grep} for an unlocker requiring a \texttt{grep} action); and (b) an LLM verifier, conditioned only on the text of \(h_t\), judges that the resulting observations entail the statement \(s_v\). Restricting the verifier to \(h_t\)---rather than letting it probe the repository on its own---ensures that establishment reflects what the trajectory has actually surfaced, not what is in principle knowable. We write \(U_t = \{v \in V_i : \mathrm{Est}(v, h_t) = 1\}\) for the established set at step \(t\). The verifier prompt is deferred to App.~\ref{app:forward_prompts}.

\textbf{Graph-aware progress.}
We want each step of a trajectory to advance \emph{coverage} of \(G_i^\star\)---establishing more of its nodes, in dependency-respecting order---without leaping to a solution-milestone move (fix plan, edit, validation) before the contextual facts it presupposes have been established. We capture both desiderata in a single per-step score \(\mathrm{Prog}_t \in [0, 1]\) that rewards new coverage and zeroes out the moment a node is established prematurely. Formally, at step \(t\), let
\[
\mathcal{A}_{t-1}(G_i^\star) \;=\; \bigl\{\,v \in V_i \setminus U_{t-1} \;:\; \mathrm{Pred}_{G_i^\star}(v) \subseteq U_{t-1}\,\bigr\}
\]
be the \emph{available frontier}: nodes legitimately discoverable from \(h_{t-1}\). 
Then we define
% Writing \(\Delta_t = U_t \setminus U_{t-1}\) for the nodes newly established at step \(t\), we call the step \emph{leaky} if any \(v \in \Delta_t\) violates \(\mathrm{Pred}_{G_i^\star}(v) \subseteq U_{t-1}\) and \emph{valid} otherwise, and define
\[
\mathrm{Prog}_t
\;=\;
\frac{|U_t \setminus U_{t-1}|}{\max(|\mathcal{A}_{t-1}(G_i^\star)|,\, 1)}
\;\cdot\;
\mathbb{1}\!\bigl[\, U_t \setminus U_{t-1} \subseteq \mathcal{A}_{t-1}(G_i^\star) \,\bigr]
\;\in\; [0, 1].
\]
% \[
% \mathrm{Prog}_t
% \;=\;
% \begin{cases}
% \,|\Delta_t| \,/\, \max(|\mathcal{A}_{t-1}|,\, 1) & \text{if step } t \text{ is valid,} \\[2pt]
% \,0 & \text{if step } t \text{ is leaky.}
% \end{cases}
% \]
The numerator counts newly established nodes; the denominator normalizes by what was eligible to be established; and the indicator hard-zeros the score if any newly established node is non-discoverable from \(h_{t-1}\). \(\mathrm{Prog}_t\) is the per-step backbone on which Phase~2 builds segment-level effectiveness \(\mathrm{Eff}_t\), by aggregating across a segment's steps and composing with a complementary groundedness gate (Sec.~\ref{sec:forward}).

\subsection{Phase 2: Receding-Horizon Bi-Objective Trajectory Realization}
\label{sec:forward}
Phase~1 supplies \(G_i^\star\) and the per-step backbone \(\mathrm{Prog}_t\); Phase~2 turns them into an executable trajectory. We grow the trajectory one segment at a time (Fig.~\ref{fig:method_overview}) and apply the bi-objective rule of Sec.~\ref{sec:problem_ours} segment by segment: at each step \(t\) we form a small pool of length-\(n\) candidate segments, commit the shortest whose effectiveness clears a local floor \(\eta_t\), and replan from the updated prefix. Both \(\mathrm{Eff}\) and \(\mathrm{Len}\) are additive over the segment partition, so under a matching floor calibration (\(\sum_t \eta_t = \eta_i\)) these greedy commits realize the trajectory-level shortest-above-floor rule on the family of rollouts the per-window pools induce (See proof sketch in App.~\ref{app:segment_optimality}).

\textbf{Candidate generation: blind seed plus single mutation.}
At each window we draw \(K\) length-\(n\) seeds from the blinded solver, \(\tilde{s}_t^{(k,0)} \sim \pi_{\mathrm{blind}}^{(n)}(\cdot \mid h_t)\), so every seed is on-prefix and on-distribution. A pure seed may miss the next available frontier node of \(G_i^\star\); the curator is therefore allowed to perturb at most one of its steps---picking a position \(t+j\) and a target \(v \in \mathcal{A}_{t+j}(G_i^\star)\), proposing a replacement response \(y'_{t+j}\) under prefix \(h_{t+j}\), and letting \(\pi_{\mathrm{blind}}\) re-roll the suffix. The window-\(t\) candidate pool \(\mathcal{S}_t\) is the union of the \(K\) pure seeds and their single-edit graph-aware variants (App.~\ref{app:candidate_generation}). Because every candidate is a blinded continuation modulo at most one localized rewrite, the pool stays close to the student distribution while remaining steerable by \(G_i^\star\), and any \((\mathrm{Eff}_t, \mathrm{Len}_t)\) gap between a seed and one of its variants is attributable to that single edit.

\textbf{Per-segment effectiveness.}
We assemble the per-segment effectiveness \(\mathrm{Eff}_t\) demanded by Sec.~\ref{sec:problem_ours} by aggregating the per-step progress over the segment, and composing with a binary groundedness gate,
\[
\mathrm{Eff}_t(s;\, G_i^\star)
\;=\;
(\sum_{\tau \in s} \mathrm{Prog}_\tau)\;\cdot\;\mathrm{Ground}_t(s),
\qquad
\mathrm{Ground}_t \in \{0, 1\}.
\]
The gate applies only to the curator-introduced edit: blinded seeds are by construction on-prefix, so we set \(\mathrm{Ground}_t(s) = 1\). A failure sends \(\mathrm{Eff}_t(s)\) to zero, removing the candidate from contention regardless of how much progress it appears to make. The binary leakage rejection inside \(\mathrm{Prog}_t\) zeros out any step in the segment with a premature establishment, while the gate inspects only the mutated step itself, so the two operate on disjoint scopes within a candidate. 

\textit{Groundedness.}\;
For a mutated step \(y\) under prefix \(h\), let \(\mathrm{Ents}(y)\) be the repository entities (file paths, identifiers, function and class names) it references and let \(\mathrm{Obs}(h)\) be the entities that have appeared in observations within \(h\); a symbolic referential-integrity check passes when \(\mathrm{Ents}(y) \subseteq \mathrm{Obs}(h)\), blocking references to an as-yet-unseen entity. A complementary LLM judge returns \(\mathrm{Claim}(y, h) \in \{0, 1\}\) on whether the reasoning of \(y\) is entailed by the observations in \(h\), ruling out premature root-cause assertions and unsupported leaps. For a mutated candidate with edit at position \(t+j\),
\[
\mathrm{Ground}_t(s)
\;=\;
\mathbb{1}\bigl[\mathrm{Ents}(y'_{t+j}) \subseteq \mathrm{Obs}(h_{t+j})\bigr]\,\cdot\,\mathrm{Claim}(y'_{t+j},\, h_{t+j}),
\]
i.e., the mutated step must clear both the symbolic check (cheap, catches concrete entity leakage) and the neural judge (catches semantic leaps that are syntactically grounded but logically unsupported). Entity-extractor patterns and the LLM-judge prompt are deferred to App.~\ref{app:forward_prompts}. 

\textbf{Selection and commit.}
Segment length is measured either as response-token mass \(\mathrm{Len}_t(s) = \sum_{y \in s} |y|\) or as step count \(\mathrm{Len}_t(s) = |s|\). 
Among all candidate segments whose effectiveness clears a local threshold \(\eta_t\), we commit the shortest (following the shortest-above-floor rule in Sec.~\ref{sec:problem_ours}),
\[
s_t^\star
\;=\;
\arg\min_{s \in \mathcal{S}_t} \mathrm{Len}_t(s)
\quad\text{s.t.}\quad
\mathrm{Eff}_t(s;\, G_i^\star) \ge \eta_t,
\]
falling back to the segment of maximum effectiveness if no candidate clears the floor. To avoid locking in a stale suffix, we adopt a half-segment stride: only the first half of \(s_t^\star\) is appended to the prefix before the procedure replans, so consecutive critic windows overlap by half their length (e.g., a window over steps \(1\text{--}10\) is followed by one over \(6\text{--}15\)). The procedure terminates when the agent submits a patch; trajectories whose patch fails the test suite \(\mathcal{T}_i\) are discarded. The end-to-end pipeline is summarized in Algorithm~\ref{alg:curation} (App.~\ref{app:algorithm}).

%% file: sections/experiments.tex
\section{Experiments}
\label{sec:experiments}

\subsection{Experimental Setup}
\label{sec:exp_setup}

\textbf{Training instances.}
We draw the training pool from SWE-Gym~\citep{pan2024swegym} ($2{,}438$ real Python issue-resolution instances over $11$ repositories), keeping the $1.8$k instances with a working Docker environment and a reference patch that passes its tests, $\mathcal{T}_i(p_i^\star)=1$ (required because \methodname consumes $p_i^\star$ as privileged information). Rejection-sampling baselines, which never inspect $p_i^\star$, run on the full $2{,}126$ executable instances. The reverse phase yields $33{,}106$ graph nodes (median $18$ per instance), dominated by static and dynamic facts; full breakdown in App.~\ref{app:node_distribution}.

\textbf{Curators, scaffold, students, baselines.}
We curate trajectories with two teachers, Qwen3-Coder-480B-A35B-Instruct~\citep{qwen3} (Qwen3-C-480B in tables) and GLM-5-FP8~\citep{glm5}, under the OpenHands~\citep{wang2024openhands} scaffold with a $100$-iteration ReAct budget. The \methodname forward phase uses a sliding window of $n=10$ steps with overlap $k=5$. We fine-tune two student backbones, Qwen2.5-Coder-14B/32B-Instruct~\citep{qwen25coder} (Qwen2.5-C-14B/32B in tables). Two prior recipes provide the baselines: \emph{Test-pass rejection sampling} (SWE-Gym)~\citep{pan2024swegym}, which keeps whole rollouts whose final patch passes $\mathcal{T}_i$; and \emph{SWE-Lego}~\citep{tao2026swelego}, which masks the SFT loss on assistant turns followed by a tool-error observation.  

\textbf{Evaluation.}
We evaluate on SWE-bench Verified ($500$ instances) and SWE-bench Lite ($300$)~\citep{jimenez2024swebench} under the same OpenHands scaffold and $100$-iteration budget. Following the bi-objective target of Sec.~\ref{sec:problem_ours}, we report two metrics per (student, teacher) cell: \emph{effectiveness} as resolve rate (Pass@1, $\uparrow$) under a single greedy rollout per instance; and \emph{efficiency} as average per-instance inference cost in US\$ ($\downarrow$), metering prompt and completion tokens at official Alibaba Cloud Model Studio list prices for Qwen2.5-Coder-14B/32B-Instruct.\footnote{Token rates from \url{https://www.alibabacloud.com/help/en/model-studio/model-pricing}; the same rate is applied to all conditions, so cost differences reflect only trajectory length.\label{fn:pricing}} Cost is averaged over the full evaluation set, capturing the inference burden a downstream user incurs whether or not the rollout resolves the issue. Full SFT hyperparameters, parallelism, and context-window extension are deferred to App.~\ref{app:training_details}.

% Define a professional dark green color for improvements (print-safe)
\definecolor{impgreen}{HTML}{006600}
% Define a macro for the delta improvements to keep spacing consistent.
% \up{x} = green ``(+x)'' for resolve-rate gains; \dn{x} = green ``($-$x)'' for cost reductions.
\newcommand{\up}[1]{\hspace{2pt}\textcolor{impgreen}{\scriptsize{(+#1)}}}
\newcommand{\dn}[1]{\hspace{2pt}\textcolor{impgreen}{\scriptsize{($-$#1)}}}
% \rg{x} = red ``(+x)'' for cost regressions over the Test-Pass RS baseline.
\newcommand{\rg}[1]{\hspace{2pt}\textcolor{red!70!black}{\scriptsize{(+#1)}}}

\subsection{Overall Effectiveness and Efficiency}
\label{sec:rq1}

We compare four trajectory-construction recipes under identical teacher, scaffold, and student-training pipelines: (i)~the test-pass rejection-sampling baseline of SWE-Gym, (ii)~the SWE-Lego process-level error-masking baseline, (iii)~\emph{\methodname (size-matched)}, in which we randomly subsample our curated trajectories down to the size of the rejection-sampled pool to control for data scale, and (iv)~\emph{\methodname (full)}, which uses every trajectory we curate from the $1.8$k-instance training pool. Following the bi-objective trajectory target of Sec.~\ref{sec:problem_ours}, Table~\ref{tab:main_results} reports both axes for each (student, teacher) pair: \emph{effectiveness} as Pass@1 resolve rate (higher is better) and \emph{efficiency} as average per-instance inference cost in US\$ (lower is better), on SWE-bench Verified and SWE-bench Lite for two student backbones (Qwen2.5-C-14B/32B) and two teacher curators (Qwen3-C-480B and GLM-5-FP8).

\begin{table*}[t]
    \centering
    \caption{\textbf{Main evaluation results on SWE-bench Verified and SWE-bench Lite.} For each (student, teacher) pair we report two metrics: \emph{effectiveness} as resolve rate (Pass@1, \%, $\uparrow$) and \emph{efficiency} as average per-instance inference cost (Cost, US\$, $\downarrow$). \textbf{Bold} denotes the best result in each row; \textcolor{impgreen}{\scriptsize{\textbf{(+X.X)}}} reports the absolute Pass@1 gain over the test-pass rejection-sampling baseline, \textcolor{impgreen}{\scriptsize{\textbf{($-$X.X)}}} the absolute cost reduction.}
    \label{tab:main_results}
    \renewcommand{\arraystretch}{1.2}
    \resizebox{\textwidth}{!}{%
    \begin{tabular}{@{} l l l c c c c c c c c @{}}
        \toprule
        \multirow{3.5}{*}{\textbf{Student}} & \multirow{3.5}{*}{\textbf{Teacher (Curator)}} & \multirow{3.5}{*}{\textbf{Metric}} & \multicolumn{4}{c}{\textbf{SWE-bench Verified}} & \multicolumn{4}{c}{\textbf{SWE-bench Lite}} \\
        \cmidrule(lr){4-7} \cmidrule(l){8-11}
        & & & \multicolumn{2}{c}{\textbf{Baselines}} & \multicolumn{2}{c}{\textbf{Ours}} & \multicolumn{2}{c}{\textbf{Baselines}} & \multicolumn{2}{c}{\textbf{Ours}} \\
        \cmidrule(lr){4-5} \cmidrule(lr){6-7} \cmidrule(lr){8-9} \cmidrule(l){10-11}
        & & & Test-Pass RS & SWE-Lego & Size-Matched & Full & Test-Pass RS & SWE-Lego & Size-Matched & Full \\
        \midrule

        % ================= 32B Model Block =================
        \multirow{4}{*}{\textbf{Qwen2.5-C-32B}}
        & \multirow{2}{*}{Qwen3-C-480B}
        & Pass@1 (\%) $\uparrow$
        & 39.6 & 40.6\up{1.0} & 42.4\up{2.8} & \cellcolor{gray!15}\textbf{50.4}\up{10.8}
        & 28.7 & 28.7\up{0.0} & 29.3\up{0.6} & \cellcolor{gray!15}\textbf{36.0}\up{7.3} \\
        & & Cost (\$) $\downarrow$
        & 0.92 & 0.95\rg{0.03} & 0.85\dn{0.07} & \cellcolor{gray!15}\textbf{0.78}\dn{0.14}
        & 0.93 & 0.93 & 0.88\dn{0.05} & \cellcolor{gray!15}\textbf{0.80}\dn{0.13} \\
        \cmidrule(l){2-11}
        & \multirow{2}{*}{GLM-5-FP8}
        & Pass@1 (\%) $\uparrow$
        & 38.4 & 38.8\up{0.4} & 39.2\up{0.8} & \cellcolor{gray!15}\textbf{49.0}\up{10.6}
        & 31.6 & 32.3\up{0.7} & 33.6\up{2.0} & \cellcolor{gray!15}\textbf{38.6}\up{7.0} \\
        & & Cost (\$) $\downarrow$
        & 0.94 & 0.93\dn{0.01} & 0.89\dn{0.05} & \cellcolor{gray!15}\textbf{0.81}\dn{0.13}
        & 0.92 & 0.95\rg{0.03} & 0.85\dn{0.07} & \cellcolor{gray!15}\textbf{0.78}\dn{0.14} \\

        \midrule

        % ================= 14B Model Block =================
        \multirow{4}{*}{\textbf{Qwen2.5-C-14B}}
        & \multirow{2}{*}{Qwen3-C-480B}
        & Pass@1 (\%) $\uparrow$
        & 36.0 & 36.2\up{0.2} & 37.6\up{1.6} & \cellcolor{gray!15}\textbf{43.2}\up{7.2}
        & 22.0 & 22.7\up{0.7} & 23.3\up{1.3} & \cellcolor{gray!15}\textbf{30.0}\up{8.0} \\
        & & Cost (\$) $\downarrow$
        & 0.92 & 0.93\rg{0.01} & 0.87\dn{0.05} & \cellcolor{gray!15}\textbf{0.78}\dn{0.14}
        & 0.94 & 0.95\rg{0.01} & 0.90\dn{0.04} & \cellcolor{gray!15}\textbf{0.83}\dn{0.11} \\
        \cmidrule(l){2-11}
        & \multirow{2}{*}{GLM-5-FP8}
        & Pass@1 (\%) $\uparrow$
        & 34.8 & 35.4\up{0.6} & 36.6\up{1.8} & \cellcolor{gray!15}\textbf{42.8}\up{8.0}
        & 24.3 & 25.3\up{1.0} & 26.3\up{2.0} & \cellcolor{gray!15}\textbf{32.0}\up{7.7} \\
        & & Cost (\$) $\downarrow$
        & 0.93 & 0.94\rg{0.01} & 0.86\dn{0.07} & \cellcolor{gray!15}\textbf{0.80}\dn{0.13}
        & 0.93 & 0.92\dn{0.01} & 0.89\dn{0.04} & \cellcolor{gray!15}\textbf{0.80}\dn{0.13} \\

        \bottomrule
    \end{tabular}%
    }
    \vspace{-0.2cm}
\end{table*}

\textbf{Results.}
Across every (student, teacher, benchmark) cell of Table~\ref{tab:main_results}, \methodname (full) is simultaneously more effective and more efficient than both baselines, lifting Pass@1 by up to $+10.8$ points (Qwen2.5-C-32B under Qwen3-C-480B on Verified) while cutting per-instance cost by $\sim\!15\%$; SWE-Lego's tool-error masking yields at most marginal Pass@1 gains and never reduces cost, since it relabels the same rejection-sampled rollouts without changing trajectory length. Two controls explain where the lift comes from. First, the size-matched configuration already beats both baselines on both axes ($+2.8$ Pass@1 and $-\$0.07$ on Verified for the 32B/Qwen3-C-480B cell), so the gain is per-trajectory \emph{quality}, not data scale. Second, moving to \methodname (full) adds another $+8.0$ Pass@1 \emph{while still lowering} cost; because the additional trajectories come from instances on which a blinded teacher rollout would have failed, this margin is supervision recovered from hard issues that rejection sampling silently discards, and the simultaneous cost drop rules out a verbosity confound. A complementary \emph{compute-matched} control (App.~\ref{app:compute_matched}) further confirms that redirecting \methodname's curation GPU-hours into $4\times$ additional teacher rollouts does not close the gap, isolating the gain from raw compute as well as from data scale.
The lift is robust: under the weaker GLM-5-FP8 teacher absolute Pass@1 gains are within $0.2$ pts and cost reductions within \$0.01 of those under Qwen3-C-480B, indicating the privileged-information factorization, not raw teacher capability, drives the improvement, and the same pattern transfers from the 32B to the 14B student.

\subsection{Trajectory Quality}
\label{sec:rq3}

The improvements in §\ref{sec:rq1} establish that \methodname jointly raises resolve rate and lowers inference cost, but they do not yet explain \emph{where} the savings come from. We trace the joint lift to a single mechanism: curated trajectories are shorter and less redundant \emph{while covering more of the fix-relevant facts} the issue presupposes, and the property is induced both in the supervision data and in the rollouts the trained student emits. Below we report the population-level shifts in length and redundancy; an end-to-end worked example on \texttt{getmoto/moto}~\#6041, tracing both phases of \methodname on this instance and contrasting the curated trajectory with the blinded rollout, is deferred to App.~\ref{app:worked_example} (Fig.~\ref{fig:wex_traj}) for space.

\begin{figure}[t]
    \centering
    \includegraphics[width=\linewidth]{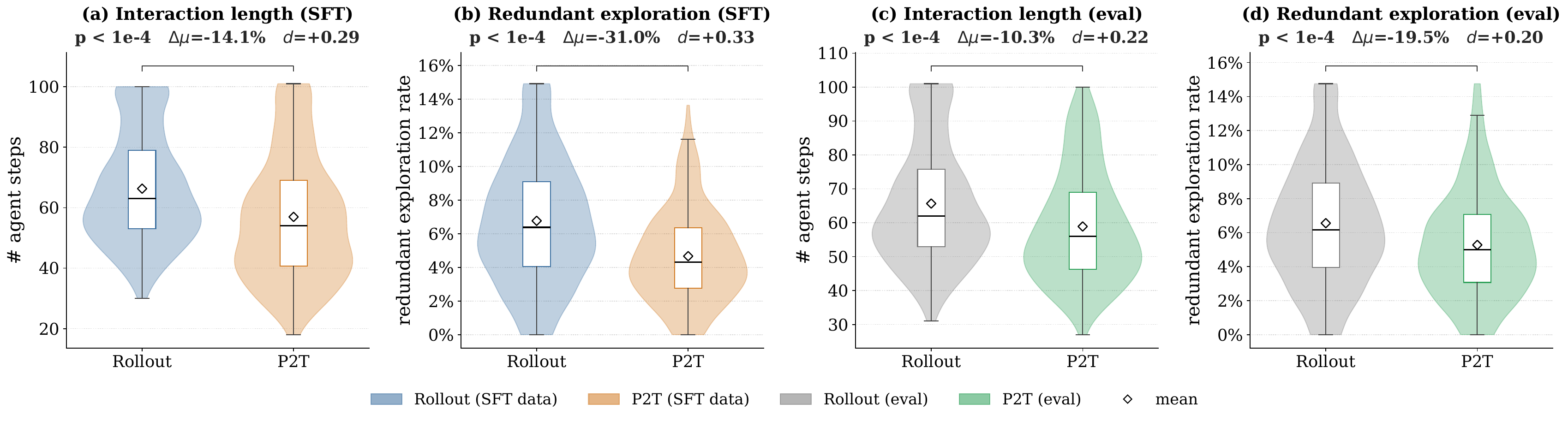}
    \caption{\textbf{Effect of \methodname on trajectory quality, traced from supervision (a, b) to evaluation (c, d).} Each panel reports the paired Mann--Whitney $p$-value, the relative shift in mean ($\Delta\mu$), and Cliff's $\delta$; diamonds mark means.}
    \label{fig:curation_effect}
\end{figure}

\textbf{Quantitative effect of curation.}
We measure two metrics: \emph{interaction length} (number of agent steps) and \emph{redundant exploration} (fraction of file-viewing actions whose visible range is fully covered by an earlier view in the same trajectory). Every comparison is restricted to instances that rollout resolve, and is measured at two pipeline stages: an \emph{SFT-data view} pairing each rejection-sampled rollout with its \methodname-curated counterpart on the same SWE-Gym instance under the GLM-5-FP8 teacher (Fig.~\ref{fig:curation_effect}\,a,b), and a \emph{student-eval view} comparing rollouts of the Qwen2.5-C-32B student trained on each supervision source on SWE-bench Verified (Fig.~\ref{fig:curation_effect}\,c,d). At the SFT stage, \methodname trajectories are $-9.3\%$ shorter ($66.3\!\to\!57$ steps) and $-31.0\%$ less redundant ($6.7\%\!\to\!4.7\%$); the heavy upper tail of rollouts that exhaust the $100$-iteration budget largely disappears (Fig.~\ref{fig:curation_effect}\,a). Both shifts are paired-Mann--Whitney significant ($p\!<\!10^{-4}$, $\delta=+0.29/+0.33$) and align with the design: the receding-horizon rule commits the shortest segment that clears the local effectiveness floor, while the groundedness critic suppresses the speculative re-inspection loops typical of unguided rollouts. Crucially, the same shifts reappear at evaluation time, where neither the curator nor any oracle is in the loop: the \methodname-trained student emits rollouts that are $-10.3\%$ shorter and $-19.5\%$ less redundant than its rejection-sampled counterpart ($p\!<\!10^{-4}$, $\delta\approx 0.20$). Behavioral cloning therefore transmits the structural property rather than merely the token-level distribution, which is the mechanism behind the $-\$0.13$--$\$0.14$ inference-cost gap in Table~\ref{tab:main_results}. 
% and because the eval-time comparison is restricted to commonly-resolved instances, the saving cannot be explained by early termination on hard issues.

\subsection{Component Ablation}
\label{sec:rq4}

We isolate the two design choices that the bi-objective program hinges on: the groundedness check that gates leakage on the effectiveness side, and the shortest-above-floor commit rule that controls length on the efficiency side. Each is removed in turn from the full pipeline; all other settings (Qwen3-C-480B teacher, OpenHands scaffold, $1.8$k-instance pool, SFT recipe) are held fixed.

\textbf{Effectiveness: groundedness check.}
Without the groundedness check, any frontier-advancing edit is committed verbatim, so curated trajectories may reference entities or claims the visible prefix does not yet support. On Qwen2.5-C-32B, removing it drops Pass@1 from $50.4\%$ to $43.2\%$ ($-7.2$ pts), with a similar $-8.2$ pt drop on Qwen2.5-C-14B. The student internalizes the same unjustified leaps at test time, exactly the failure mode the check was designed to block.

\textbf{Efficiency: shortest-above-floor commit.}
Replacing the shortest-above-floor rule with a uniform-random pick from the candidate pool of equally effective segments removes the only term that pressures the trajectory to be short. Curated trajectories grow accordingly: average step count rises from $72.5$ to $77.8$ ($+7.3\%$) and per-trajectory token length from $64$k to $72$k ($+12.5\%$). Pass@1 also slips from $50.4\%$ to $48.8\%$ on Qwen2.5-C-32B, indicating that the shorter trajectories are not just cheaper but carry less imitation noise: behavioral cloning amplifies the lower-information steps that the rule would have pruned. Together, the two ablations show that the bi-objective rule is load-bearing on both axes simultaneously rather than trading them off.

\textbf{Is $G^\star$ doing the work?}
A value-of-information study (App.~\ref{app:voi}) confirms that the gains track $G^\star$ itself, not the forward-phase scaffolding: progressively disclosing $G^\star$ to a blinded reference solver lifts its Pass@1 from $29$--$35\%$ to $94$--$97\%$ under both teachers, and incidental $G^\star$ coverage of unguided rollouts correlates strongly with success ($r\!\approx\!0.36$). Both indicate that $G^\star$ encodes substantive prerequisites rather than post-hoc narration of $p^\star$.

%% file: sections/conclusion.tex
\section{Conclusion}
\label{sec:conclusion}

We presented \methodname, a framework that converts a reference patch into per-step process supervision for software-engineering agents without ever exposing the patch in the trajectory shown to the student. The pipeline factorizes curation into a reverse decomposition that distills a sufficient yet non-leaky prerequisite graph $G^\star$ from $p^\star$, and a forward grounded realization that uncovers $G^\star$ through ordinary tool calls under a hybrid groundedness critic and a surprisal trust region. On SWE-bench Verified, training on $1.8$k curated SWE-Gym instances improves Qwen2.5-Coder-32B/14B-Instruct by $+10.8$/$+7.2$ points Pass@1 over an outcome-filtered baseline, with consistent gains on SWE-bench Lite and across two structurally different teacher curators. Controlled subsampling, a value-of-information study with progressive disclosure of $G^\star$, an observational coverage analysis, a quantitative trajectory-quality comparison, and a component ablation jointly indicate that the gains are attributable to the prerequisite-graph factorization itself rather than to data scale or any single safeguard. We discuss limitations and outlook in App.~\ref{app:limitations}.

%% file: sections/appendix.tex
\newpage
% \section{Appendix}
\addtocontents{toc}{\protect\setcounter{tocdepth}{3}}
\renewcommand{\contentsname}{Appendix}

\hypersetup{linkcolor=black}
\tableofcontents %
\hypersetup{linkcolor=red}
\newpage

\appendix

\section{Technical appendices and supplementary material}
This appendix provides material complementary to the main paper: prerequisite-graph statistics (App.~\ref{app:node_distribution}), a value-of-information study for $G^\star$ (App.~\ref{app:voi}), reverse-phase agent prompts (App.~\ref{app:reverse_prompts}), the segment-wise commit analysis (App.~\ref{app:segment_optimality}), candidate generation in detail (App.~\ref{app:candidate_generation}), the full curation algorithm (App.~\ref{app:algorithm}), forward-phase agent prompts (App.~\ref{app:forward_prompts}), an end-to-end worked example (App.~\ref{app:worked_example}), training details (App.~\ref{app:training_details}), the compute-matched RS baseline (App.~\ref{app:compute_matched}), and limitations and outlook (App.~\ref{app:limitations}).

\section{Prerequisite-graph node distribution}
\label{app:node_distribution}

\begin{figure}[h]
    \centering
    \includegraphics[width=\linewidth]{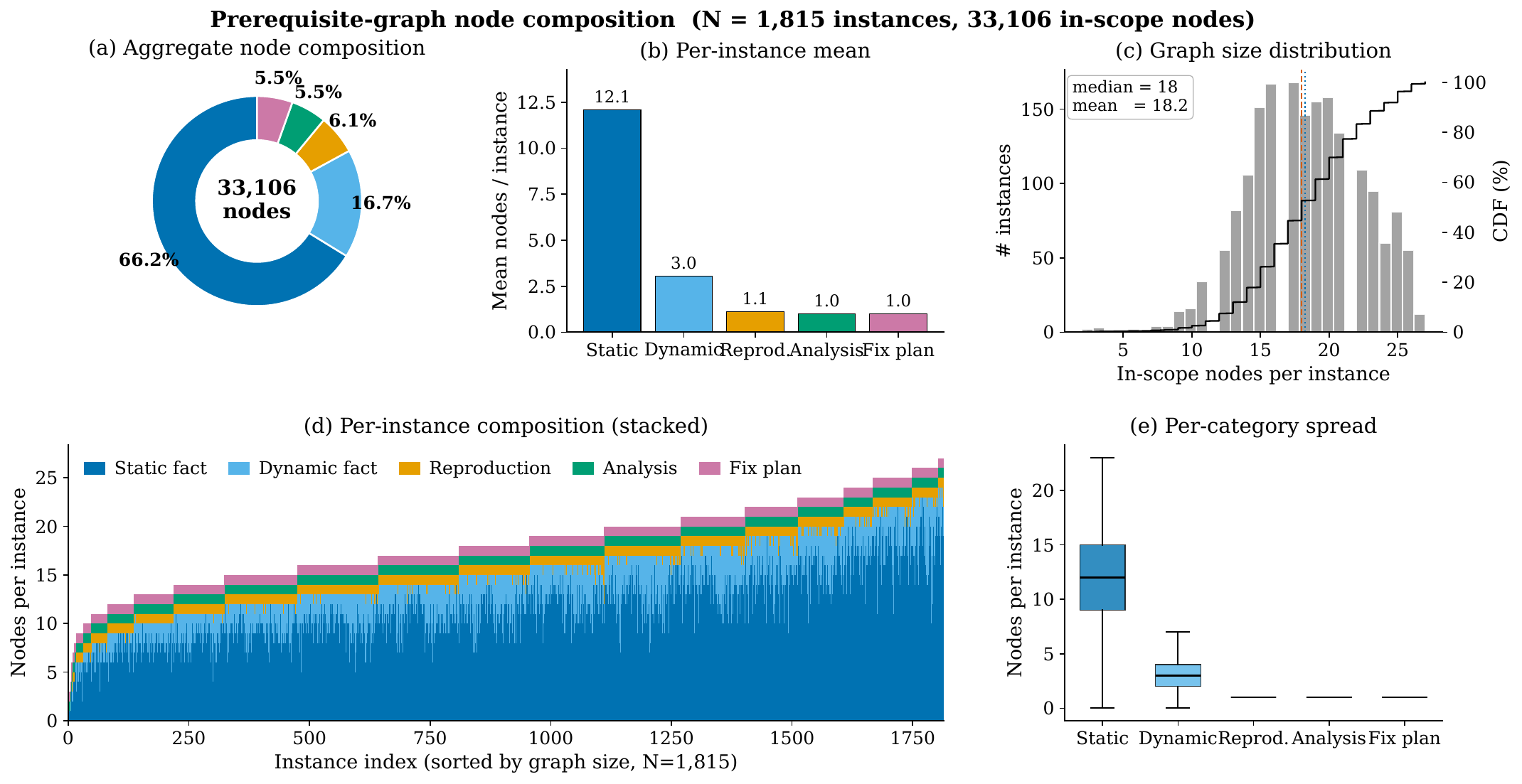}
    \caption{\textbf{Composition of the prerequisite graphs $\{G_i^\star\}$ across the $N\!=\!1{,}815$ SWE-Gym training instances ($33{,}106$ in-scope nodes).} \textbf{(a)}~Aggregate node-type breakdown: static facts dominate ($66.2\%$), with dynamic facts ($16.7\%$) and the three artifact categories (reproduction, analysis, fix plan) at $5$--$6\%$ each. \textbf{(b)}~Per-instance mean count by category (bars) and fraction of instances containing at least one node of that category (right axis): every instance has at least one static fact and fix plan; dynamic facts are sparser on average but still present in most instances. \textbf{(c)}~Distribution of total in-scope graph size across instances (median $18$, mean $18.2$) and its CDF. \textbf{(d)}~Per-instance stacked composition with instances sorted by graph size, showing that static-fact share grows with size while artifact counts remain stable. \textbf{(e)}~Per-category spread: artifact categories are tightly concentrated, whereas static and dynamic facts account for nearly all of the cross-instance variance.}
    \label{fig:node_distribution}
\end{figure}

Figure~\ref{fig:node_distribution} characterizes the prerequisite graphs $\{G_i^\star\}$ produced by the reverse phase on the $N\!=\!1{,}815$ SWE-Gym training instances used to curate \methodname trajectories, comprising $33{,}106$ in-scope nodes in total. Panel~(a) shows that the aggregate node population is dominated by \emph{static facts} ($66.2\%$, e.g., file/symbol locations and type signatures readable from the repo at $h$), followed by \emph{dynamic facts} ($16.7\%$, observations that require executing code), and the three artifact categories---reproduction, analysis, and fix plan---each contributing roughly $5$--$6\%$. Panel~(b) reports the per-instance mean count alongside the fraction of instances that contain at least one node of each category: every instance has at least one static fact and a fix plan ($100\%$ coverage), reproduction and analysis artifacts appear in essentially every instance as well, while dynamic facts have lower per-instance mass ($3.0$ nodes on average) but appear whenever the issue's behavior is meaningfully runtime-dependent. Panel~(c) gives the distribution of total in-scope graph size: the mode is concentrated around $15$--$20$ nodes (median $18$, mean $18.2$), with a thin right tail extending past $25$. Panels~(d,e) examine variation across instances. The stacked composition in (d), with instances sorted by total graph size, shows that the static-fact share grows roughly linearly with graph size, while artifact counts ($\le\!2$ per category in nearly all instances) are remarkably stable; the per-category spreads in (e) confirm this---artifact categories are tightly concentrated near their median, whereas static and dynamic facts carry essentially all of the cross-instance variance.

Three consequences for our method follow. First, because static facts dominate, the bulk of the curator's frontier-advancement work consists of grounded read-only inspection actions (file reads, symbol lookups, ripgrep), which are cheap and naturally on-policy for a non-privileged solver. Second, the small but stable artifact budget per instance ($\sim\!1$--$3$ nodes each for reproduction/analysis/plan) bounds the number of synthesized intermediate steps the forward phase has to inject, keeping curated trajectories close in length to organic rollouts. Third, the long tail in panel~(c) identifies a sub-population of instances with large graphs ($>\!25$ nodes) where the gap between curated and rejection-sampled trajectories is largest, since these are precisely the instances on which a single greedy blinded rollout is least likely to incidentally cover the required prerequisites (cf.\ App.~\ref{app:voi}).

\section{Value of information in the prerequisite graph}
\label{app:voi}

The improvements in Sec.~\ref{sec:rq1} establish that \methodname trajectories help \emph{a student}, but they do not on their own show that the prerequisite graph $G^\star$ itself carries the right information; one might worry that the gains come entirely from the forward-phase machinery (segment-level bi-objective optimization selection, groundedness gating) and that $G^\star$ is little more than a post-hoc rationalization of $p^\star$. This appendix tests the alternative directly through two complementary protocols. First, an \emph{interventional} disclosure study progressively reveals nodes of $G^\star$ to a blinded reference solver and measures the resulting end-to-end resolve rate. Second, an \emph{observational} coverage study verifies that, even when $G^\star$ is never exposed, blinded rollouts whose incidental coverage of $G^\star$ is highest are also the ones most likely to succeed. Both directions converge on the same conclusion.

\paragraph{Protocol (interventional disclosure).}
We define a nested chain of information bundles
\(
B^{(0)} \subset B^{(1)} \subset B^{(2)} \subset B^{(3)} \subset B^{(4)},
\)
each augmenting the visible prefix at the start of an episode:
\begin{itemize}
    \item $B^{(0)}$: issue $x$ only;
    \item $B^{(1)}$: $B^{(0)} +$ all facts in $V_F^\star$ (context);
    \item $B^{(2)}$: $B^{(1)} +$ the reproduction-script artifact;
    \item $B^{(3)}$: $B^{(2)} +$ the root-cause analysis artifact;
    \item $B^{(4)}$: $B^{(3)} +$ the fix-plan artifact.
\end{itemize}
Crucially, $p^\star$ is never disclosed at any stage, so each bundle is something a non-privileged solver could in principle have constructed for itself. We restrict the study to the $1.8$k training instances on which $p_i^\star$ passes $\mathcal{T}_i$ (the same pool used for our curated trajectories) so that variation in resolve rate is attributable to the bundle, not to broken evaluation environments. For each bundle and each teacher, the same teacher model is then used as the reference solver under the OpenHands scaffold with the standard $100$-iteration budget, and we measure Pass@1 averaged over a single rollout per instance.

\begin{figure}[h]
    \centering
    \begin{minipage}[b]{0.52\linewidth}
        \centering
        \includegraphics[width=\linewidth]{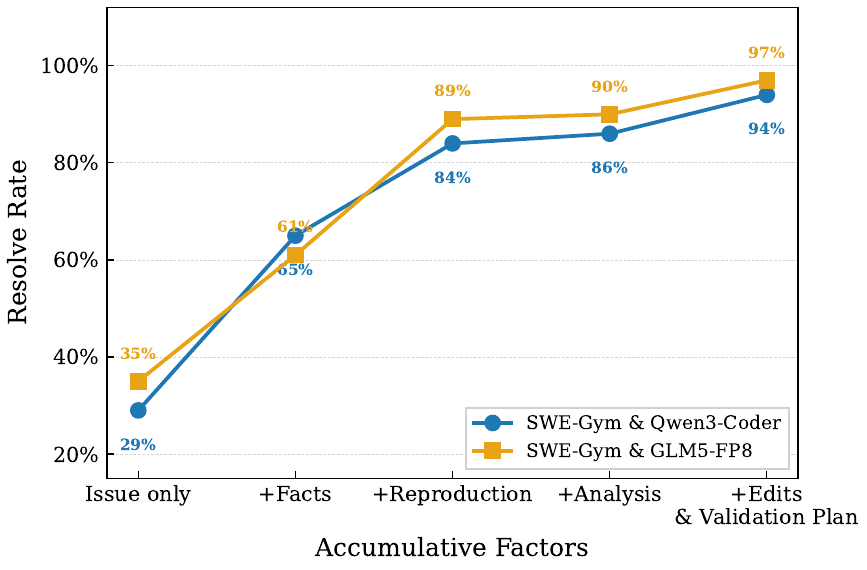}
        \subcaption{Progressive disclosure of $G^\star$ to a blinded solver.}
        \label{fig:voi_disclose}
    \end{minipage}\hfill
    \begin{minipage}[b]{0.46\linewidth}
        \centering
        \includegraphics[width=\linewidth]{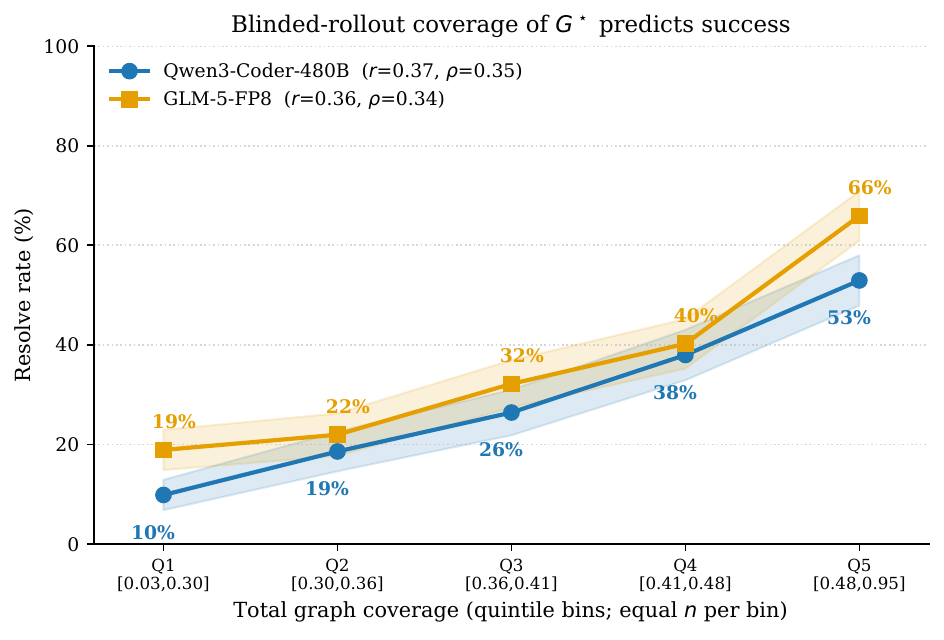}
        \subcaption{Observational coverage of $G^\star$ in blinded rollouts vs.\ resolve rate.}
        \label{fig:voi_cov}
    \end{minipage}
    \caption{\textbf{Value of information in $G^\star$.} \textbf{(a)}~Resolve rate of a blinded reference solver on the $1.8$k training instances as elements of the prerequisite graph are progressively revealed; the oracle patch $p^\star$ is never exposed. Each addition contributes a non-negative marginal gain under both teachers, with facts and the reproduction script accounting for most of the lift. \textbf{(b)}~When the graph is \emph{not} disclosed, instances on which a blinded rollout incidentally covers more nodes of $G^\star$ are resolved at substantially higher rates: rollouts in the top coverage quintile succeed at $53$--$66\%$, vs.\ $10$--$19\%$ in the bottom quintile, with a consistent positive correlation under both teachers ($r\!\approx\!0.36$, $\rho\!\approx\!0.35$).}
    \label{fig:voi}
\end{figure}

\paragraph{Results (interventional).}
Figure~\ref{fig:voi} shows that the resolve rate increases monotonically as more of $G^\star$ is revealed, under both teachers. From issue-only context, the reference solver resolves $29\%$ ($35\%$) of instances under the Qwen3-Coder-480B (GLM-5-FP8) curator. Adding the fact statements alone---no scripts, no plans---more than doubles this to $65\%$ ($61\%$). Revealing the reproduction script delivers the largest single jump, to $84\%$ ($89\%$); the root-cause analysis adds a further $+2$ ($+1$) points; and the fix plan plus validation stub take performance to $94\%$ ($97\%$). Every marginal addition is non-negative for both teachers.

Three observations follow. First, the magnitude of the lift from $B^{(0)}$ to $B^{(4)}$ in Fig.~\ref{fig:voi_disclose}---roughly $+65$ percentage points---is far too large to be explained away as cosmetic narration: the curated graph encodes substantive, action-relevant knowledge. Second, the bulk of the improvement comes from facts ($+\!\!\sim\!\!30$ pts) and the reproduction script ($+\!\!\sim\!\!20$ pts), confirming that the reverse phase's two staples---atomic fact distillation and concrete artifact scaffolding---are precisely the components that carry information for a non-privileged solver. Third, the consistency of the trend across two structurally different teachers indicates that the prerequisite-graph structure, rather than any teacher-specific stylistic bias, is what supplies the value of information.

\paragraph{Observational coverage predicts success without disclosure.}
A natural concern with the disclosure protocol is that prepending structured text to the prompt may help for reasons unrelated to the \emph{content} of $G^\star$---e.g., framing or anchoring effects. We therefore complement Fig.~\ref{fig:voi_disclose} with an \emph{observational} test in which the graph is never exposed to the solver. For each of the $1.8$k instances, we take a fully blinded rollout under each teacher (issue only, no $G^\star$, no $p^\star$) and apply the same unlock criterion the curator uses during forward realization---a node is counted as covered when an unlocker action (or an equivalent action) is executed and the associated statement (or an equivalent concept) is encoded in the trajectory. The total coverage of an instance is then the fraction of nodes in $V_F^\star \cup V_A^\star$ that are covered. We bin instances into quintiles of equal size by this coverage and report the resolve rate per bin in Fig.~\ref{fig:voi_cov}.

Resolve rate increases monotonically with coverage under both teachers, climbing from $10\%$/$19\%$ in the lowest-coverage quintile to $53\%$/$66\%$ in the highest, with Pearson $r\!\approx\!0.36$ and Spearman $\rho\!\approx\!0.35$. Because the solver is given no privileged information in this analysis, the correlation cannot be an artifact of prompt augmentation: it shows that the very prerequisites identified by reverse decomposition are, on the issues a blinded solver \emph{does} resolve, the ones it tends to establish on its own. The two panels of Fig.~\ref{fig:voi} thus triangulate the same conclusion from opposite directions---disclosing $G^\star$ helps, and independently establishing $G^\star$ correlates with success---providing converging evidence that $G^\star$ tracks information genuinely required to solve the issue rather than post-hoc narration of $p^\star$, and justifying the use of $G^\star$ as the per-step effectiveness signal in Sec.~\ref{sec:forward}.

\section{Reverse-phase agent prompts and implementation details}
\label{app:reverse_prompts}

We instantiate the proposer--critic loop of Sec.~\ref{sec:reverse} with two LLM agents that share a tool-mediated view of the repository (file-viewer, ripgrep, sandboxed \texttt{python}/\texttt{pytest}) and exchange JSON node sets. Both agents see the issue \(d_i\), the repository \(R_i\), the reference patch \(p_i^\star\), and the test suite \(\mathcal{T}_i\); only the proposer is allowed to introduce new nodes, and only the critic is allowed to delete or revise them. The DAG-organization step is a deterministic post-pass: an edge is drawn from \(u\) to \(v\) iff \(v\)'s unlocker action references an entity (file, symbol, line range, runtime value) first surfaced by \(u\)'s unlocker observation, or \(v\) is an artifact whose type strictly follows \(u\)'s in the canonical order \(\text{repro}\!\to\!\text{analysis}\!\to\!\text{plan}\!\to\!\text{edit}\!\to\!\text{validation}\).

\paragraph{Unlocker taxonomy.}
Each node carries an \emph{unlocker} \(u_v=(\textsc{action},\,\textsc{observation})\) drawn from a fixed taxonomy: (i)~\textsc{view}\(\langle\text{file},\text{lines}\rangle\) for static reads; (ii)~\textsc{view}\,\texttt{problem\_statement} for issue-derived facts; (iii)~\textsc{bash}\(\langle\text{command}\rangle\) for grep, runtime probing, or test execution (dynamic facts); (iv)~\textsc{create}\(\langle\text{path},\text{content}\rangle\) and \textsc{str\_replace}\(\langle\text{file},\text{old},\text{new}\rangle\) for reproduction scripts and edits; (v)~\textsc{think}\(\langle\text{text}\rangle\) for analysis and fix-plan artifacts. Every action string must be copy-pasteable; abbreviations such as \texttt{...} are rejected by the critic on sight.

\paragraph{Stopping criterion.}
Writing \(V^{(t)}\) for the node set after critic round \(t\), the loop terminates when (a) \(V^{(t)}=V^{(t-1)}\) (set-level fixed point under hashed canonical statements), or (b) the proposer returns \(\Delta V=\varnothing\) for two consecutive rounds, or (c) a hard cap of \(T_{\max}=6\) rounds is reached. In our runs, \(96\%\) of instances converge by round~3.

\paragraph{LLM and decoding.}
Both agents are run on the same teacher backbone used for blinded rollouts (Qwen3-Coder-480B or GLM-5-FP8) with temperature \(0.4\), top-\(p\) \(0.95\), max output \(8\)k tokens, JSON-mode constrained decoding, and a \(40\)-step tool-call budget per round. Outputs are validated against a JSON schema; on schema violation the agent is re-prompted up to twice with the validator error before the round is dropped.

The two prompt templates below are \emph{abstracted} versions of the production prompts: we preserve the role, inputs, hard rules, and output schema verbatim but elide engineering boilerplate (tool-API specifications, output-format examples, retry instructions, and repository-specific stop-lists) for readability. 

\begin{tcolorbox}[colback=gray!4, colframe=gray!50, sharp corners, boxsep=2pt,
                  left=4pt, right=4pt, top=2pt, bottom=2pt,
                  fontupper=\scriptsize, breakable,
                  title=\textbf{Proposer prompt (reverse phase, Sec.~\ref{sec:reverse})}]
\textbf{Role.} You are the \emph{Proposer}. Working backwards from a known-correct reference patch \(p^\star\), your job is to enumerate the contextual facts and solution milestones a non-privileged developer would need to discover before \(p^\star\) becomes derivable. You are NOT producing a description of \(p^\star\) --- you are producing the prerequisite knowledge that motivates \(p^\star\).

\textbf{Inputs.} Issue \(d_i\); repository \(R_i\) (read/grep/exec via tools); reference patch \(p_i^\star\); failing-to-passing tests; current node set \(V^{(t)}\) with critic feedback \(\phi^{(t)}\).

\textbf{Task.} Emit \(\Delta V^{(t+1)}\): a JSON list of new candidate nodes that close the largest remaining gap to \(p^\star\). Each node has \texttt{id}, \texttt{node\_type} \(\in\{\text{fact},\text{reproduce\_script},\text{issue\_analysis},\text{fix\_plan},\text{code\_edit},\text{validation}\}\), a one-claim \texttt{statement}, and a fully-specified \texttt{unlocker}. Cover \emph{every} logically distinct change in \(p^\star\): imports, registration entries, exception classes, parameter additions, helper reuse, schema/template updates.

\textbf{Hard rules.}\;
(R1) One claim per fact; split compounds.\;
(R2) The unlocker action is restricted to the taxonomy above; no \texttt{[view] golden patch}, \texttt{[view] hints}, or any reference to \(p^\star\).\;
(R3) Action strings are complete and replayable --- no \texttt{...}, no prose placeholders.\;
(R4) Observations are the actual tool output you obtained when you executed the action --- do not paraphrase from \(p^\star\).\;
(R5) For every \texttt{reproduce\_script} and \texttt{validation} node, you must execute the script in the sandbox and record \texttt{output\_before\_fix}/\texttt{actual\_output}; nodes without recorded execution are rejected.\;
(R6) For every \texttt{code\_edit}, \texttt{old\_str} must match the current pre-fix source byte-for-byte.

\textbf{Investigation protocol.}\; (i)~Index every distinct change in \(p^\star\). (ii)~For each change, trace the call chain backwards to the user-facing symptom and forwards to the user-facing effect. (iii)~Search the repository for \emph{analogous implementations} and cite at least one concrete location (file + line range) when \(p^\star\) follows an existing pattern. (iv)~Probe runtime behavior with \texttt{python -c} or \texttt{pytest} when types, dispatch, or values are not statically obvious. (v)~Narrow the bug: test the failing case AND the working case to confirm the fix scope is minimal. (vi)~Apply the critic feedback \(\phi^{(t)}\) and address each cited gap.

\textbf{Output.} A JSON object \(\{\text{instance\_id},\, \text{nodes}: [\,\ldots\,]\}\) written via \texttt{create\_file} to the designated path. Nothing else.
\end{tcolorbox}

\begin{tcolorbox}[colback=gray!4, colframe=gray!50, sharp corners, boxsep=2pt,
                  left=4pt, right=4pt, top=2pt, bottom=2pt,
                  fontupper=\scriptsize, breakable,
                  title=\textbf{Critic prompt (reverse phase, Sec.~\ref{sec:reverse})}]
\textbf{Role.} You are the \emph{Critic}. Your job is to enforce desideratum~(ii), \emph{non-leakage}: every surviving node's unlocker must be conceivable from the issue, the repository, and the node's predecessors --- never from \(p^\star\). You also enforce minimality and discoverability.

\textbf{Inputs.} \(d_i,\, R_i,\, p_i^\star,\, V^{(t)}\cup\Delta V\) (the proposer's latest output).

\textbf{Per-node verdict.} For each candidate \(v\), assign exactly one of:
\textbf{keep} (passes all checks); \textbf{prune} (fails at least one hard check); \textbf{revise} (salvageable with the noted edit). Output a JSON list with \texttt{id}, \texttt{verdict}, \texttt{reasons[]}, and (for revise) \texttt{patched\_node}.

\textbf{Hard checks (any failure $\Rightarrow$ prune).}\;
(H1) \textsc{Patch leakage.} The statement describes what \(p^\star\) does (e.g.\ ``the patch adds X'', ``the fix is to call Y'') rather than a property of the pre-fix codebase or runtime. \;
(H2) \textsc{Forbidden unlocker.} Action references \(p^\star\), the test patch, or hidden hints; uses a tool outside the taxonomy; or contains an abbreviation that prevents replay.\;
(H3) \textsc{Unverified observation.} The recorded \texttt{observation} cannot be reproduced by re-executing \texttt{action} on the pre-fix repository.\;
(H4) \textsc{Premature artifact.} A \texttt{code\_edit} node appears without an upstream \texttt{fix\_plan}; a \texttt{fix\_plan} appears without an upstream \texttt{issue\_analysis}; a \texttt{validation} appears without the \texttt{code\_edit} it validates.\;
(H5) \textsc{Compound claim.} A fact bundles two or more distinct propositions.\;
(H6) \textsc{Edit drift.} A \texttt{code\_edit} node's \texttt{old\_str} does not match the current source byte-for-byte, or the union of all edits does not reproduce \(p^\star\).

\textbf{Soft checks (issue \texttt{revise}).}\;
(S1) Two facts state the same claim with different evidence \(\Rightarrow\) merge, keeping the stronger evidence.\;
(S2) An action is replayable but unnecessarily wide (e.g.\ \texttt{view 1--500}) when 10 lines suffice \(\Rightarrow\) tighten the line range.\;
(S3) A static fact whose statement obviously requires execution to verify \(\Rightarrow\) re-tag as dynamic and provide a probe.

\textbf{Feedback to proposer.} For every cited gap, append a one-sentence \texttt{$\phi$}-entry naming (a) the missing fact category, and (b) a concrete next investigation (e.g.\ ``add a fact identifying the analogous handler in the same file''). The feedback is consumed verbatim by the next proposer round.

\textbf{Output.} A JSON object \(\{\text{instance\_id},\, \text{verdicts}: [\,\ldots\,],\, \text{feedback}: [\,\ldots\,]\}\) written via \texttt{create\_file}. The next proposer round receives \(V^{(t+1)} = V^{(t)} \cup \{v : \text{verdict}(v)=\text{keep}\} \cup \{\text{patched}(v) : \text{verdict}(v)=\text{revise}\}\) together with the feedback list.
\end{tcolorbox}

\section{Segment-wise commits and the trajectory-level objective}
\label{app:segment_optimality}

Sec.~\ref{sec:forward} commits trajectories one segment at a time: at each window \(t\) it forms a candidate pool \(\mathcal{S}_t\), restricts to the locally non-dominated set \(\mathcal{P}_t = \mathrm{ND}(\mathcal{S}_t)\) under \((\mathrm{Eff}_t, \mathrm{Len}_t)\), and commits
\(
s_t^\star = \arg\min_{s \in \mathcal{P}_t} \mathrm{Len}_t(s) \;\text{s.t.}\; \mathrm{Eff}_t(s; G_i^\star) \ge \eta_t.
\)
Section~\ref{sec:problem_ours}, by contrast, formulates a \emph{trajectory-level} rule: among trajectories satisfying \(\mathrm{Eff}_i(\tau) \ge \eta_i\), pick the shortest. We sketch why the segment-wise commits realize the trajectory-level rule on the family of trajectories that the per-window pools induce.

\paragraph{Setup.}
A trajectory built by the procedure decomposes into disjoint windows, \(\tau = s_1 \oplus \cdots \oplus s_W\), where \(\oplus\) denotes concatenation along the prefix. Both objectives are additive over this partition:
\[
\mathrm{Len}_i(\tau) \;=\; \sum_{t=1}^{W} \mathrm{Len}_t(s_t),
\qquad
\mathrm{Eff}_i(\tau) \;=\; \sum_{t=1}^{W} \mathrm{Eff}_t(s_t),
\]
where \(\mathrm{Len}\) is response-token mass (or step count) and \(\mathrm{Eff}_t(s_t) = \sum_{\tau \in s_t} \mathrm{Prog}_\tau\) on segments that pass the groundedness gate. Additivity holds because windows are non-overlapping and \(\mathrm{Prog}_\tau\) is computed against the realized-node set \(U_{\tau-1}\) at the step's own prefix, with the leakage rejection inside \(\mathrm{Prog}_\tau\) and the groundedness gate \(\mathrm{Ground}_t\) operating on disjoint scopes within a segment (Sec.~\ref{sec:forward}).

\paragraph{Floor calibration.}
Let the per-window floors be calibrated so that \(\sum_{t=1}^{W} \eta_t = \eta_i\). In practice we set a constant \(\eta_t = \eta_i / W^\star\) for an expected horizon \(W^\star\) and absorb stochasticity in the fallback rule of Sec.~\ref{sec:forward} (commit the maximum-effectiveness segment when no candidate clears the floor); the proof statement below pertains to the deterministic case in which the floor is met at every window.

\paragraph{Greedy feasible set.}
Conditioned on the prefix produced by previous commits, each window induces a candidate pool \(\mathcal{S}_t\) and an admissible subset
\(
\mathcal{F}_t = \{ s \in \mathcal{P}_t \mid \mathrm{Eff}_t(s) \ge \eta_t \}.
\)
The \emph{greedy feasible family} is
\[
\mathcal{T}_{\mathrm{greedy}} \;=\; \bigl\{\, s_1 \oplus \cdots \oplus s_W \,:\, s_t \in \mathcal{F}_t \text{ under prefix } s_1 \oplus \cdots \oplus s_{t-1} \,\bigr\},
\]
i.e.\ all trajectories formable by picking one admissible segment per window from the pools the procedure actually encounters.

\paragraph{Claim.}
The trajectory \(\tau^\star = s_1^\star \oplus \cdots \oplus s_W^\star\) returned by the segment-wise procedure satisfies
\[
\mathrm{Eff}_i(\tau^\star) \;\ge\; \eta_i,
\qquad
\mathrm{Len}_i(\tau^\star) \;=\; \min_{\tau \in \mathcal{T}_{\mathrm{greedy}}} \mathrm{Len}_i(\tau),
\]
and is therefore the shortest-above-floor trajectory in \(\mathcal{T}_{\mathrm{greedy}}\).

\paragraph{Proof sketch.}
\emph{(i) Floor.} For each window \(t\), the commit rule enforces \(\mathrm{Eff}_t(s_t^\star) \ge \eta_t\). Summing across windows and using additivity of \(\mathrm{Eff}\) gives
\(
\mathrm{Eff}_i(\tau^\star) = \sum_t \mathrm{Eff}_t(s_t^\star) \ge \sum_t \eta_t = \eta_i.
\)
\emph{(ii) Length-optimality.} Fix any \(\tau = s_1 \oplus \cdots \oplus s_W \in \mathcal{T}_{\mathrm{greedy}}\). Pointwise minimality of \(s_t^\star\) within \(\mathcal{F}_t\) gives \(\mathrm{Len}_t(s_t^\star) \le \mathrm{Len}_t(s_t)\) at every \(t\); summing and using additivity of \(\mathrm{Len}\) yields \(\mathrm{Len}_i(\tau^\star) \le \mathrm{Len}_i(\tau)\). Combined with (i), \(\tau^\star\) is feasible and length-minimal in \(\mathcal{T}_{\mathrm{greedy}}\), which is the trajectory-level shortest-above-floor rule restricted to that family.

\paragraph{Scope.}
The proof shows \emph{no Pareto regret within \(\mathcal{T}_{\mathrm{greedy}}\)}, not global optimality over the full trajectory space \(\mathcal{T}(\mathcal{I}_i)\) of Sec.~\ref{sec:problem_ours}: greedy commits change the prefix, and a different early commit could open downstream pools containing strictly better segments. Two choices control this gap. First, replanning (the receding-horizon \(\delta < n\) execution in Sec.~\ref{sec:forward}) shrinks the prefix that any single commit locks in, reducing the loss from greediness. Second, drawing seeds from \(\pi_{\mathrm{blind}}\) and bounding curator intervention to one edit per segment keeps the per-window pools well-distributed across reasonable continuations, so the greedy family is a representative slice of the trajectory space rather than a degenerate corner of it. The empirical comparison against rejection sampling and the ablations in Sec.~\ref{sec:experiments} measure how much this restriction costs in practice.

\section{Candidate generation in detail}
\label{app:candidate_generation}

This appendix expands the candidate-pool construction sketched in Sec.~\ref{sec:forward}.

\paragraph{Blinded seeds.}
At window \(t\) we draw \(K\) length-\(n\) continuations
\[
\tilde{s}_t^{(k,0)} \sim \pi_{\mathrm{blind}}^{(n)}(\cdot \mid h_t),
\qquad k = 1, \dots, K,
\]
each executed in a sandbox copy of the environment so that side effects on \(R_i\) (file edits, test runs, package installs) do not leak across candidates.

\paragraph{Single-mutation variants.}
For each seed \(\tilde{s}_t^{(k,0)}\) the curator may additionally pick a position \(t+j\) (\(0 \le j < n\)) and a target node \(v \in \mathcal{A}_{t+j}(G_i^\star)\) available under the simulated prefix \(h_{t+j}^{(k)}\), then draw a one-step replacement response
\[
y'_{t+j} \sim q_{\mathrm{mut}}(\cdot \mid h_{t+j}^{(k)},\, v)
\]
that points the next action at \(v\)'s unlocker. The sandbox is rolled back to step \(t+j\), \(y'_{t+j}\) is executed, and the remaining \(n - j\) steps of the suffix are regenerated from \(\pi_{\mathrm{blind}}\) under the new prefix. Indexing the resulting variants by their target \(m\), the window-\(t\) pool is
\[
\mathcal{S}_t \;=\; \{\tilde{s}_t^{(k,0)}\}_{k=1}^K \;\cup\; \{s_t^{(k,m)}\}_{k,m}.
\]

\paragraph{Why one edit per segment.}
Confining curator intervention to a single step per segment plays two roles. First, it isolates causality inside the local Pareto problem: every variant differs from some pure seed by exactly one localized rewrite, so any \((\mathrm{Eff}_t, \mathrm{Len}_t)\) gap between them can be attributed to that edit, which makes the per-segment selection rule of Sec.~\ref{sec:forward} well-posed and the groundedness gate cheap to evaluate (it inspects only \(y'_{t+j}\)). Second, it keeps each committed action a blinded action up to a single rewrite, bounding the trajectory's deviation from the student-facing distribution and ensuring the resulting tokens remain safe imitation targets under SFT.

\section{Full curation algorithm}
\label{app:algorithm}

Algorithm~\ref{alg:curation} gives the end-to-end pseudocode for \methodname, combining the process graph distillation of Sec.~\ref{sec:reverse} with the receding-horizon bi-objective trajectory realization of Sec.~\ref{sec:forward}. The notation matches Sec.~\ref{sec:method}: \(G_i^\star=(V_i,E_i)\) is the distilled process graph, \(U_t\) is the realized-node set, \(\mathcal{A}_t\) the available frontier, \(\mathrm{Prog}_\tau\) the per-step progress score, \(\mathrm{Ground}_t\) the binary groundedness gate, and \((\eta_t, n, K, \delta)\) the per-window floor, segment length, seed count, and commit horizon.

\begin{algorithm}[htbp]
\caption{\methodname: oracle-guided process graph distillation and trajectory realization}
\label{alg:curation}
\begin{algorithmic}[1]
\Require Task instance \(\mathcal{I}_i=(d_i, R_i, E_i, \mathcal{T}_i, p_i^\star)\); blinded solver \(\pi_{\mathrm{blind}}\); window length \(n\); seeds per window \(K\); commit horizon \(\delta \le n\); per-window floor \(\eta_t\)
\Ensure Curated trajectory \(\tau_i^\star\) admitted into \(\mathcal{D}_{\mathrm{ours}}\), or \(\bot\)

\Statex \textbf{Phase 1: Process Graph Distillation (Sec.~\ref{sec:reverse})}
\State \(V \leftarrow \varnothing\)
\Repeat \Comment{proposer--critic loop}
    \State \(\Delta V \leftarrow \mathrm{Proposer}(d_i, R_i, V, p_i^\star)\) \Comment{(i) sufficiency: close logical gap to \(p_i^\star\)}
    \State \(V \leftarrow \mathrm{Critic}(V \cup \Delta V, d_i, R_i)\) \Comment{(ii) non-leakage: prune nodes whose unlocker presupposes \(p_i^\star\)}
\Until{\(V\) stabilizes}
\State \(E \leftarrow \{(u,v) : u \text{ must be realized before } v\text{'s unlocker applies}\}\) \Comment{(iii) feasible ordering}
\State \(G_i^\star \leftarrow (V, E)\)

\Statex \textbf{Phase 2: Receding-Horizon Bi-Objective Realization (Sec.~\ref{sec:forward})}
\State \(h \leftarrow d_i\); \;\(U \leftarrow \varnothing\); \;\(\tau \leftarrow ()\); \;\(t \leftarrow 0\)
\While{agent has not call finish tool}
    \State \(\mathcal{A} \leftarrow \{v \in V \setminus U : \mathrm{Pred}_{G_i^\star}(v) \subseteq U\}\) \Comment{available frontier}
    \State \(\mathcal{S}_t \leftarrow \varnothing\)
    \For{\(k = 1, \dots, K\)} \Comment{$K$ blinded seeds, sandboxed}
        \State \(\tilde{s}^{(k,0)} \sim \pi_{\mathrm{blind}}^{(n)}(\cdot \mid h)\); \;add to \(\mathcal{S}_t\) with \(\mathrm{Ground}=1\)
        \For{each position \(j \in \{0,\dots,n{-}1\}\) and target \(v \in \mathcal{A}_{t+j}(G_i^\star)\) under \(\tilde{s}^{(k,0)}\)}
            \State Roll back sandbox to step \(t+j\); draw \(y'_{t+j} \sim q_{\mathrm{mut}}(\cdot \mid h_{t+j}^{(k)}, v)\) \Comment{single-mutation variant}
            \State Execute \(y'_{t+j}\); regenerate suffix from \(\pi_{\mathrm{blind}}\) \(\Rightarrow s^{(k,v)}\)
            \State \(\mathrm{Ground}_t(s^{(k,v)}) \leftarrow \mathbb{1}[\mathrm{Ents}(y'_{t+j}) \subseteq \mathrm{Obs}(h_{t+j})] \cdot \mathrm{Claim}(y'_{t+j}, h_{t+j})\)
            \State Add \(s^{(k,v)}\) to \(\mathcal{S}_t\)
        \EndFor
    \EndFor
    \For{each \(s \in \mathcal{S}_t\)} \Comment{score under \((\mathrm{Eff}_t, \mathrm{Len}_t)\)}
        \State Simulate \(s\) on \(h\); for each step \(\tau \in s\) compute \(U_\tau, \mathcal{A}_{\tau-1}, \Delta_\tau = U_\tau \setminus U_{\tau-1}\)
        \State \(\mathrm{Prog}_\tau \leftarrow \dfrac{|\Delta_\tau|}{\max(|\mathcal{A}_{\tau-1}|,\, 1)} \cdot \mathbb{1}\!\bigl[\,\Delta_\tau \subseteq \mathcal{A}_{\tau-1}\,\bigr]\) \Comment{indicator hard-zeros leaky steps}
        \State \(\mathrm{Eff}_t(s) \leftarrow \mathrm{Ground}_t(s) \cdot \sum_{\tau \in s} \mathrm{Prog}_\tau\); \;\(\mathrm{Len}_t(s) \leftarrow \sum_{y \in s} |y|\) \;(or \(|s|\))
    \EndFor
    \State \(\mathcal{P}_t \leftarrow \mathrm{ND}(\mathcal{S}_t)\) under \((\mathrm{Eff}_t, \mathrm{Len}_t)\); \;\(\mathcal{F}_t \leftarrow \{s \in \mathcal{P}_t : \mathrm{Eff}_t(s) \ge \eta_t\}\)
    \If{\(\mathcal{F}_t \ne \varnothing\)}
        \State \(s_t^\star \leftarrow \arg\min_{s \in \mathcal{F}_t} \mathrm{Len}_t(s)\) \Comment{shortest-above-floor}
    \Else
        \State \(s_t^\star \leftarrow \arg\max_{s \in \mathcal{P}_t} \mathrm{Eff}_t(s)\) \Comment{fallback}
    \EndIf
    \State \(s_t^{\mathrm{commit}} \leftarrow\) first \(\delta\) steps of \(s_t^\star\)
    \State Execute \(s_t^{\mathrm{commit}}\) on the live environment; append \((y, o)\) pairs to \(\tau\) and \(h\); update \(U\)
    \State \(t \leftarrow t + \delta\)
\EndWhile
\State \(\hat{p} \leftarrow \mathrm{patch}(\tau)\)
\If{\(\mathcal{T}_i(\hat{p}) = 1\)} \Return \(\tau_i^\star \leftarrow \tau\) \Else{} \Return \(\bot\) \EndIf
\end{algorithmic}
\end{algorithm}

\section{Forward-phase agent prompts and implementation details}
\label{app:forward_prompts}

The forward phase (Sec.~\ref{sec:forward}) instantiates four specialized components, each with a fixed prompt template: a \emph{curator} that proposes single-step mutations toward an available frontier node; a symbolic \emph{entity extractor} that implements the referential-integrity half of \(\mathrm{Ground}_t\); a neural \emph{claim-grounding judge} that implements the entailment half of \(\mathrm{Ground}_t\); and a node-\emph{establishment verifier} that decides \(\mathrm{Est}(v,h_t)\) for the per-step progress score \(\mathrm{Prog}_t\) (Sec.~\ref{sec:reverse}). All four components see only the visible prefix \(h_t\), the issue \(d_i\), and (for the curator and verifier) the relevant graph nodes; none of them ever sees \(p_i^\star\) or the test suite. The prompt panels below are \emph{abstracted} versions of the production prompts --- role, inputs, hard rules, and output schema are preserved verbatim, while tool-API specifications, output-format examples, and engineering boilerplate are elided for readability. The full production prompts will be released with the code. The curator additionally receives the available frontier \(\mathcal{A}_{t+j}\) and the realized set \(U_{t+j}\) at the position it chooses to mutate.

\paragraph{Sliding-window and hyperparameter settings.}
We use a window length \(n=10\) ReAct steps, half-segment stride (commit horizon \(\delta=5\), so consecutive windows overlap by \(5\) steps), and \(K=2\) blinded seeds per window plus up to one mutation per seed (\(\le 4\) candidates per pool). The per-window floor is \(\eta_t = \max(1,\,|\mathcal{A}_{t-1}|)\cdot 0.5\) (i.e., realize at least half the currently available frontier), with the fallback rule of Sec.~\ref{sec:forward} when no candidate clears the floor. Seeds are drawn from \(\pi_{\mathrm{blind}}\) at temperature \(0.6\) (top-\(p\) \(0.95\), max output \(2\)k tokens per step), with a per-seed sandbox snapshot rolled back after scoring. The curator and judge run on the same teacher backbone as the seeds at temperature \(0.2\); the entity extractor is a deterministic Python pass over the regex set listed below.

\paragraph{Entity-extractor pattern set (symbolic check).}
\(\mathrm{Ents}(y)\) and \(\mathrm{Obs}(h)\) are defined as the union of strings matched by the following Python-regex family, applied to the assistant message and tool-call arguments of \(y\) (resp.\ to all observations and assistant messages in \(h\)):

\begin{tcolorbox}[colback=gray!4, colframe=gray!50, sharp corners, boxsep=2pt,
                  left=4pt, right=4pt, top=2pt, bottom=2pt,
                  fontupper=\scriptsize\ttfamily, breakable,
                  title=\textbf{Entity regex set (case-sensitive; flags re.MULTILINE)}]
FILE\_PATH\_REL    \;:=\; (?:[\textbackslash{}w.\textbackslash{}-]+/)+[\textbackslash{}w.\textbackslash{}-]+\textbackslash{}.(?:py|pyx|pyi|c|cpp|h|js|ts|json|yaml|yml|toml|cfg|md|txt|sh)\\
FILE\_PATH\_ABS    \;:=\; /(?:workspace|testbed|opt|usr|home)/[\textbackslash{}w./\textbackslash{}-]+\\
DOTTED\_MODULE     \;:=\; (?:[a-z\_][a-z0-9\_]*\textbackslash{}.){1,}[a-zA-Z\_][a-zA-Z0-9\_]*\\
QUALIFIED\_NAME    \;:=\; \textbackslash{}b[A-Z][A-Za-z0-9\_]*(?:\textbackslash{}.[A-Za-z\_][A-Za-z0-9\_]*)+\textbackslash{}b\\
IDENTIFIER\_DEF   \;:=\; (?:def|class|async\textbackslash{} def)\textbackslash{}s+([A-Za-z\_][A-Za-z0-9\_]*)\\
IDENTIFIER\_REF   \;:=\; \textbackslash{}b[\_a-zA-Z][\_a-zA-Z0-9]\{2,\}\textbackslash{}b\quad (filtered by stop-list of Python keywords + 200 most common English words)\\
LINE\_REF         \;:=\; (?:line|lines|L)\textbackslash{}s\#?\textbackslash{}d+(?:\textbackslash{}s*[-\textendash]\textbackslash{}s*\textbackslash{}d+)?\\
ERROR\_TYPE       \;:=\; \textbackslash{}b[A-Z][A-Za-z]*(?:Error|Exception|Warning)\textbackslash{}b\\
SHELL\_FLAG       \;:=\; (?<=\textbackslash{}s)--?[a-zA-Z][a-zA-Z\_\textbackslash{}-]+\\
NUMERIC\_LITERAL  \;:=\; \textbackslash{}b\textbackslash{}d\{3,\}\textbackslash{}b\quad (e.g., issue numbers, large constants; small ints are excluded as non-discriminative)
\end{tcolorbox}

The symbolic check passes iff every match in \(y\) is also a match in \(h\), modulo a path-normalization step that strips workspace prefixes and a trailing-suffix collapse for nested attribute access (\texttt{a.b.c} matches if either \texttt{a.b.c} or \texttt{b.c} appears in \(h\)). Single-token English identifiers shorter than three characters and members of the keyword stop-list are excluded from \(\mathrm{Ents}(y)\) to avoid spurious failures. Empirically the symbolic gate fires on \(13.4\%\) of mutated candidates, and \(91\%\) of those are downstream rejected by the neural judge as well.

\paragraph{Curator prompt.}
Given a sandboxed seed \(\tilde s_t^{(k,0)}\), the curator picks a position \(j\) and a target \(v\in\mathcal{A}_{t+j}\), then writes a one-step replacement \(y'_{t+j}\) intended to advance toward \(v\)'s unlocker. It does not receive \(p^\star\); it sees only the issue, the visible prefix at position \(t+j\), and the natural-language statement and unlocker action of the targeted graph node.

\begin{tcolorbox}[colback=blue!3, colframe=blue!40, sharp corners, boxsep=2pt,
                  left=4pt, right=4pt, top=2pt, bottom=2pt,
                  fontupper=\scriptsize, breakable,
                  title=\textbf{Curator prompt (forward phase, Sec.~\ref{sec:forward})}]
\textbf{Role.} You are guiding a blinded coding agent. The agent's next ReAct step is currently \texttt{candidate}, but you believe a small redirection would unlock more useful evidence. You may rewrite \emph{at most one} step.

\textbf{Visible to you.} (a) The issue description \(d_i\). (b) The full prefix \(h_{t+j}\) including all prior assistant messages, tool calls, and observations. (c) The target node \(v\): a single statement and its unlocker action (e.g., ``read \texttt{responses/security\_groups.py} lines 183--197 to compare with sibling \texttt{describe\_security\_groups}''). (d) The candidate step \texttt{candidate} the blinded agent originally produced.

\textbf{Hidden from you.} The reference patch, the test suite, the rest of the graph, and any future steps. Do not speculate about them.

\textbf{Task.} Produce \(y'_{t+j}=(c'_{t+j},\,a'_{t+j})\), a single ReAct response whose action moves the trajectory toward \(v\)'s unlocker.

\textbf{Hard rules.}\;
(C1) The reasoning \(c'_{t+j}\) must be entailed by what is visible in \(h_{t+j}\) --- no claim that requires evidence the prefix has not yet produced. Frame the redirection as a natural next thought (``before going deeper into X, let me first check Y'').\;
(C2) Every entity (file path, identifier, line number) you mention must already appear somewhere in \(h_{t+j}\), either via the issue text or via earlier observations. If \(v\)'s unlocker mentions an entity not yet observed, choose a smaller approach action that surfaces that entity first.\;
(C3) The action \(a'_{t+j}\) is a single tool call from the agent's tool ontology (\texttt{str\_replace\_editor view}, \texttt{execute\_bash}, \texttt{think}, \texttt{create\_file}, \texttt{str\_replace}); arguments must be fully specified.\;
(C4) Do not write or refer to any meta-concept (``oracle'', ``graph'', ``target node'', ``reference patch''). Your output must read as if a skilled developer wrote it from scratch.\;
(C5) Stay within one step. Do not stack multiple tool calls; do not pre-announce future steps in detail.

\textbf{Output.} JSON: \(\{\text{response\_content}: c'_{t+j},\, \text{response\_tool\_calls}: [a'_{t+j}]\}\). Nothing else.
\end{tcolorbox}

\paragraph{Claim-grounding judge prompt.}
The judge implements \(\mathrm{Claim}(y'_{t+j},h_{t+j})\in\{0,1\}\) and is invoked only on the curator-mutated step. It runs after the symbolic check and is by construction blinded to \(p^\star\) and to the target node.

\begin{tcolorbox}[colback=red!3, colframe=red!40, sharp corners, boxsep=2pt,
                  left=4pt, right=4pt, top=2pt, bottom=2pt,
                  fontupper=\scriptsize, breakable,
                  title=\textbf{Claim-grounding judge prompt (forward phase, Sec.~\ref{sec:forward})}]
\textbf{Role.} You are an independent reviewer. You decide whether a single proposed assistant step is \emph{grounded} in the prior interaction history --- meaning every claim it makes and every entity it references is either stated in the issue, visible in earlier tool outputs, or trivially derivable from them.

\textbf{Inputs.}\; (i) Issue \(d_i\). (ii) Full untruncated prefix \(h_{t+j}\) (every assistant message and every observation). (iii) Proposed step \(y'_{t+j}\) (its reasoning text and its tool call).

\textbf{Hidden.} You do not have access to the reference patch, the test suite, or any oracle. Judge solely from (i)--(iii).

\textbf{Decision criteria (all must hold).}\;
(J1) \textsc{Reasoning entailment.} Every assertion in the reasoning text is either present in \(d_i\) / \(h_{t+j}\) verbatim, or is an immediate one-step inference from content that is. Disallowed: identifying a root cause whose symptom has not yet been observed; asserting that a method ``probably'' does X without having read its body.\;
(J2) \textsc{Argument reachability.} Every argument of the tool call (file paths, line numbers, symbol names, shell flags) appears in \(h_{t+j}\) or \(d_i\), or is a transparent transformation thereof (e.g.\ widening a previously viewed line range by a few lines).\;
(J3) \textsc{No oracle artifacts.} The text contains no reference to ``oracle'', ``golden patch'', ``reference solution'', ``ground truth'', or any meta-concept about being guided.\;
(J4) \textsc{Continuity.} The step is a natural incremental continuation of \(h_{t+j}\); there is no sudden jump in knowledge or unexplained change of focus.

\textbf{Output.} JSON: \(\{\text{valid}: \text{true}\,|\,\text{false},\ \text{reasons}: [\ldots]\}\), where \texttt{reasons} cites the violated criterion and the offending span when \texttt{valid}=false. Return \texttt{valid}=true iff (J1)--(J4) all hold.
\end{tcolorbox}

\paragraph{Establishment verifier prompt.}
For each candidate window the verifier is invoked on every \((v,h_\tau)\) pair where \(v\) is a node not yet in \(U_{\tau-1}\) and \(h_\tau\) extends \(h_{\tau-1}\) by exactly one step. It returns \(\mathrm{Est}(v,h_\tau)\in\{0,1\}\), feeding directly into \(U_\tau\), \(\mathcal{A}_{\tau-1}\), and \(\mathrm{Prog}_\tau\) (Sec.~\ref{sec:reverse}).

\begin{tcolorbox}[colback=green!3, colframe=green!40!black!50, sharp corners, boxsep=2pt,
                  left=4pt, right=4pt, top=2pt, bottom=2pt,
                  fontupper=\scriptsize, breakable,
                  title=\textbf{Establishment verifier prompt (Sec.~\ref{sec:reverse})}]
\textbf{Role.} Decide whether the trajectory prefix \(h_\tau\) has \emph{established} the graph node \(v=(s_v,\eta_v,u_v)\), where \(s_v\) is a natural-language statement and \(u_v\) is a required interaction (e.g., a file view, a grep, a script execution).

\textbf{Two-part criterion.} Establishment requires both:\;
(E1) \textsc{Action match.} Some action in \(h_\tau\) matches the requirement specified by \(u_v\). Equivalence is judged by intent, not by string match: a \texttt{view} of a wider line range that includes the target lines counts; an \texttt{execute\_bash python -c '\ldots'} that exercises the same code path as a probe script counts; a \texttt{grep} over a superset of the requested directory counts.\;
(E2) \textsc{Statement entailment.} The observation produced by the matching action, conditioned only on \(h_\tau\), entails \(s_v\). You may NOT consult the repository directly --- restrict the judgment to what the trajectory has actually surfaced. If the observation is consistent with \(s_v\) but does not entail it (e.g., a function name appears but its body is not shown), return false.

\textbf{Inputs.} \(d_i\); the full prefix \(h_\tau\); the node \(v\). You do NOT see \(p^\star\), the test suite, or other graph nodes.

\textbf{Output.} JSON: \(\{\text{established}: \text{true}\,|\,\text{false},\ \text{matched\_action}: \text{(step index or null)},\ \text{evidence}: \text{(short quote from the matching observation)},\ \text{reason}: \text{(one sentence; required when false)}\}\).
\end{tcolorbox}

\paragraph{Termination, retries, and trajectory acceptance.}
A candidate that fails either half of \(\mathrm{Ground}_t\) is dropped from \(\mathcal{S}_t\) without re-prompting --- the curator may not retry on the same \((j,v)\) pair within a window, since unbounded retries would let the curator search for a leakage-free phrasing of an inherently leaky claim. If all mutated candidates fail and no pure seed clears \(\eta_t\), the fallback rule of Sec.~\ref{sec:forward} commits the maximum-effectiveness pure seed. The forward loop terminates when the agent emits the \texttt{finish} action; the resulting patch is then run against \(\mathcal{T}_i\) and the trajectory is admitted into \(\mathcal{D}_{\mathrm{ours}}\) iff all targeted tests pass.

\section{End-to-end worked example}
\label{app:worked_example}

This appendix traces a single SWE-Gym instance,
\texttt{getmoto/moto\#6041} (\emph{``ec2.describe\_security\_group\_rules does
not use filter''}), end-to-end through the \methodname pipeline. The instance
is representative because (i) the oracle patch is one line, giving a clean
contrast between a wandering blinded rollout and the curated trajectory;
(ii) the blinded teacher over-engineers the fix on the model layer and breaks
an existing test, so curation has to both \emph{remove} a long detour and
\emph{insert} a small correction; and (iii) the distilled graph $G^\star$
contains every node category (static fact, dynamic fact, reproduction,
analysis, plan, edit, validation), so a single instance illustrates the full
recipe.

\subsection{Issue $x$ and oracle patch $p^\star$}
\label{app:wex_issue}

\begin{tcolorbox}[colback=gray!4, colframe=gray!50, sharp corners, boxsep=2pt,
                  left=4pt, right=4pt, top=2pt, bottom=2pt,
                  fontupper=\small, title=\textbf{Issue $x$}]
\texttt{ec2.describe\_security\_group\_rules} does not use filter.
Calling it with \texttt{Filters=[\{"Name":"group-id","Values":[sg\_id]\}]}
returns rules for \emph{all} security groups (default + dummy) instead of
only the dummy SG. The user's repro uses \texttt{boto3 + moto.mock\_ec2} and
expects exactly the default egress rule for the new SG; actually gets two.
\end{tcolorbox}

\begin{tcolorbox}[colback=gray!4, colframe=gray!50, sharp corners, boxsep=2pt,
                  left=4pt, right=4pt, top=2pt, bottom=2pt,
                  fontupper=\scriptsize\ttfamily,
                  title=\textbf{Oracle patch $p^\star$ (one line)}]
--- a/moto/ec2/responses/security\_groups.py\\
+++ b/moto/ec2/responses/security\_groups.py\\
@@ class SecurityGroups(EC2BaseResponse):\\
\phantom{XX}def describe\_security\_group\_rules(self) -> str:\\
\phantom{XXXX}group\_id = self.\_get\_param("GroupId")\\
{\color{red!70!black}-\phantom{XX}filters = self.\_get\_param("Filter")}\\
{\color{green!50!black}+\phantom{XX}filters = self.\_filters\_from\_querystring()}\\
\phantom{XXXX}rules = self.ec2\_backend.describe\_security\_group\_rules(group\_id, filters)
\end{tcolorbox}

\noindent The fix replaces a flat \texttt{\_get\_param("Filter")} (which looks
for a literal \texttt{"Filter"} key in the querystring, finds none, and
returns \texttt{None}) with the EC2 helper used by every other
\texttt{describe\_*} handler in the same file.

\subsection{Phase 1 --- Process graph distillation}
\label{app:wex_reverse}

\begin{figure}[h]
\centering
\includegraphics[width=\linewidth]{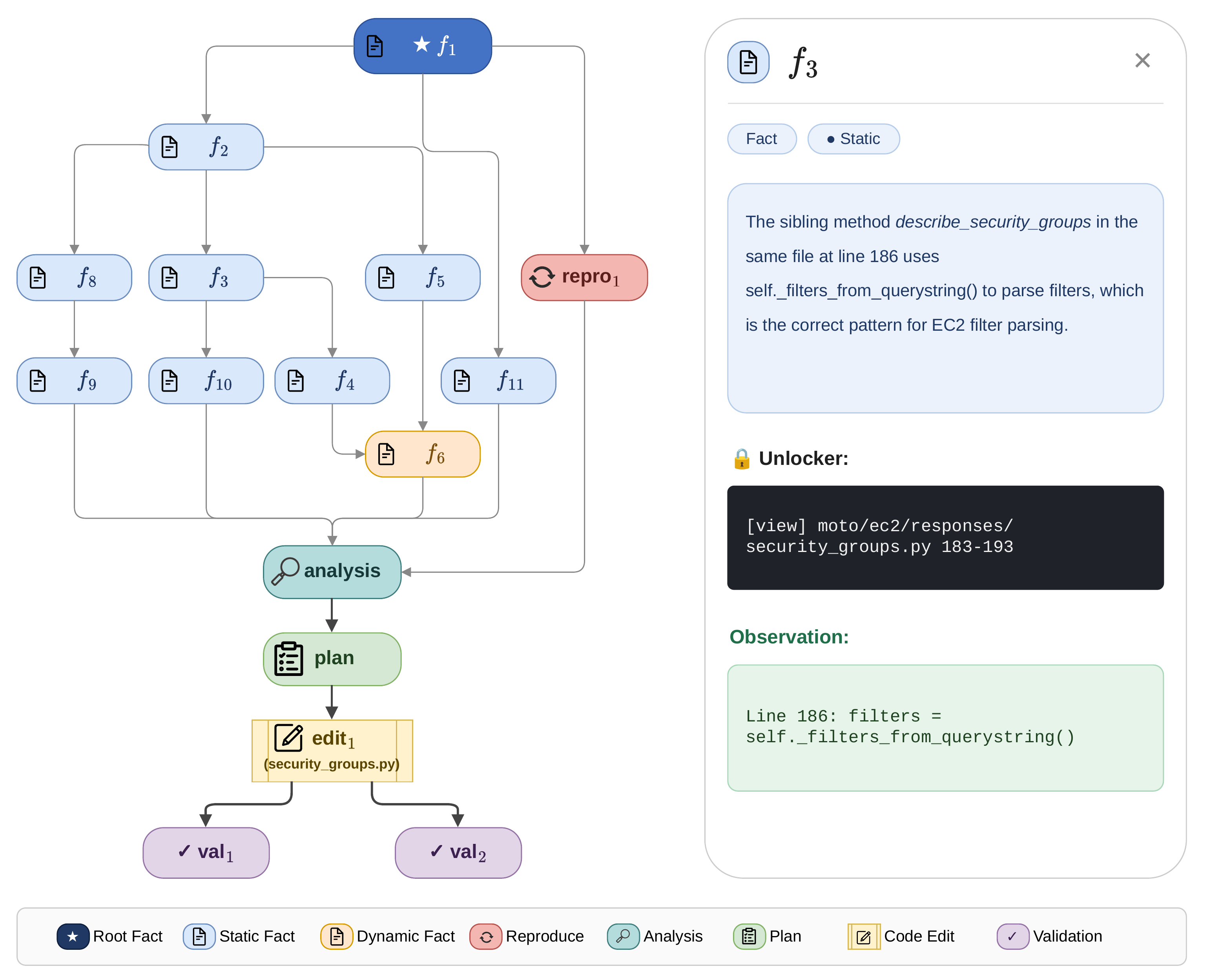}
\caption{Distilled prerequisite graph $G^\star$ for \texttt{moto\#6041}.
Static facts (blue) form the contextual layer; the dynamic fact $f_6$
(orange) requires execution; the artifact layer (green/grey/red/purple)
encodes the reproduction, analysis, plan, edit, and validation milestones.
Edges denote prerequisite relations enforced by the critic during
distillation.}
\label{fig:wex_graph}
\end{figure}

The reverse phase converts $p^\star$ into a latent process graph
$G^\star=(V,E)$ with \textbf{10 contextual-fact nodes} and \textbf{6
solution-milestone nodes} (1 reproduction, 1 analysis, 1 plan, 1 edit, 2
validations). The converged fact set $V_F^\star$ is summarised in
Table~\ref{tab:wex_facts} and the full DAG is shown in
Fig.~\ref{fig:wex_graph}.

\begin{table}[h]
\centering\scriptsize
\caption{Converged fact set $V_F^\star$ for \texttt{moto\#6041}. Each node
carries a natural-language statement, a type tag, and an explicit unlocker
(omitted from this table for brevity; see Fig.~\ref{fig:wex_graph}).}
\label{tab:wex_facts}
\begin{tabular}{@{}llp{0.74\linewidth}@{}}
\toprule
id & type & statement \\ \midrule
$f_1$ & static  & Repro shows \texttt{describe\_security\_group\_rules} returning rules for \emph{all} SGs when a \texttt{group-id} filter is passed. \\
$f_2$ & static  & Line~197 of the response handler uses \texttt{self.\_get\_param('Filter')} to parse filters. \\
$f_3$ & static  & Sibling \texttt{describe\_security\_groups} (line~186) uses \texttt{self.\_filters\_from\_querystring()} -- the standard EC2 pattern. \\
$f_4$ & static  & \texttt{\_filters\_from\_querystring} in \texttt{EC2BaseResponse} parses the numbered \texttt{Filter.N.*} querystring into a \texttt{\{name: values\}} dict. \\
$f_5$ & static  & \texttt{\_get\_param(p)} does an \emph{exact} key lookup; for \texttt{p='Filter'} it never matches \texttt{Filter.1.Name} and returns \texttt{None}. \\
$f_6$ & dynamic & At runtime the querystring contains \texttt{Filter.1.Name}/\texttt{Filter.1.Value.1}; \texttt{\_get\_param('Filter')$\to$None}, helper $\to$ correct dict. \\
$f_8$ & static  & Backend \texttt{describe\_security\_group\_rules} delegates to \texttt{describe\_security\_groups(group\_ids, filters)}. \\
$f_9$ & static  & Backend filter step (line~540: \texttt{if filters:}) is \emph{skipped} when \texttt{filters} is \texttt{None}, so all groups match. \\
$f_{10}$ & static & \texttt{\_filters\_from\_querystring} is the standard pattern across \emph{all} other EC2 \texttt{describe\_*} handlers (\texttt{instances.py}, \texttt{hosts.py}, \dots). \\
$f_{11}$ & static & The existing test masks the bug -- it iterates \emph{all} returned rules looking for an id match, never asserts the count. \\
\bottomrule
\end{tabular}
\end{table}

\paragraph{One iteration of the proposer--critic loop.}
We illustrate non-leakage enforcement on a single round. Starting from
$V^{(0)}=\{f_1\}$, the Proposer reads $p^\star$ and proposes
$\Delta V_F^{(1)}=\{c_1,\ldots,c_5\}$ (Table~\ref{tab:wex_proposer}).

\begin{table}[h]
\centering\scriptsize
\caption{Proposer round~1 candidates for \texttt{moto\#6041}.}
\label{tab:wex_proposer}
\begin{tabular}{@{}lp{0.55\linewidth}p{0.30\linewidth}@{}}
\toprule
cand. & proposed statement & critic verdict \\ \midrule
$c_1$ & Buggy line~197 uses \texttt{\_get\_param('Filter')}. & keep ($\to f_2$) \\
$c_2$ & \emph{The fix is to call \texttt{\_filters\_from\_querystring()} at line~197.} & \cellcolor{red!10}\textbf{prune (leaks $p^\star$)} \\
$c_3$ & Sibling \texttt{describe\_security\_groups} uses the correct helper. & keep ($\to f_3$) \\
$c_4$ & At runtime the querystring uses \texttt{Filter.N.*} keys, not \texttt{Filter}. & keep ($\to f_6$) \\
$c_5$ & Backend silently skips filtering when \texttt{filters is None}. & keep ($\to f_9$) \\
\bottomrule
\end{tabular}
\end{table}

\noindent The critic's feedback to the Proposer is
\begin{quote}\itshape\small
$\phi$: ``$c_2$ is the patch itself; its unlocker presupposes knowledge of
$p^\star$. Decompose into (a) a fact identifying the bug location, (b) a
fact identifying the correct alternative pattern observed elsewhere in the
repo, and (c) a plan node that conjoins them.''
\end{quote}
A second iteration adds $\Delta V_F^{(2)}=\{f_4,f_5,f_8,f_{10},f_{11}\}$ to
close residual gaps (the helper definition, the \texttt{\_get\_param}
semantics, the backend delegation, the cross-file prevalence of the pattern,
and the masking test); the loop converges. A subsequent
\textsc{ScaffoldArtifactDAG} pass appends
$V_A^\star=\{\text{repro}_1,\text{analysis},\text{plan},\text{edit}_1,\text{val}_1,\text{val}_2\}$
and links every node into the DAG of Fig.~\ref{fig:wex_graph}.

\subsection{Phase 2 --- Receding-horizon trajectory realization}
\label{app:wex_forward}

We trace one sliding window in detail. Setup: window length $n=10$, commit
horizon $\delta=4$, $K=2$ blinded seeds plus up to one graph-aware mutation
each ($\le 4$ candidates per window), per-window floor $\eta_t$ calibrated
so that $\sum_t\eta_t\!\approx\!|V|$.

\paragraph{State at window start.}
The window we trace begins at prefix end-step $t=22$ of the original blinded
rollout. The realized-node set and available frontier are
\[
U_{22} = \{f_1,\,\text{repro}_1,\,f_2\},
\qquad
\mathcal{A}_{22} = \{f_3,\,f_5,\,f_8,\,f_{11}\}
\]
($f_3,f_5,f_8$ have $f_2\in U_{22}$ as their only predecessor; $f_{11}$ has $f_1\in U_{22}$).
This window is exactly where the blinded teacher takes its wrong turn: it
omits $f_3$ (the sibling-method comparison) and is therefore biased to
``fix'' the backend rather than the response handler.

\paragraph{Two blinded seeds.}
Both seeds are length-10 continuations from $\pi_{\mathrm{blind}}(\cdot\mid h_{22})$.

\begin{table}[h]
\centering\scriptsize
\caption{Window-22 candidate pool ($n{=}10$ steps each). $\mathrm{Eff}_{22}=
\sum_\tau \mathrm{Prog}_\tau$ aggregates per-step ratios
$\mathrm{Prog}_\tau=|\Delta_\tau|/|\mathcal{A}_{\tau-1}|\in[0,1]$;
$\mathrm{Len}_{22}$ is response-token mass. Both mutated variants tie at
$\mathrm{Eff}_{22}=0.70$ above the floor $\eta_{22}\!\approx\!0.5$;
$s^{(0,f_3)}$ wins the length tie-break and is committed as $s_{22}^\star$.}
\label{tab:wex_pool}
\begin{tabular}{@{}llcccc@{}}
\toprule
candidate & $\Delta$ realized & $|\Delta|$ & $\mathrm{Ground}$ & $\mathrm{Eff}_{22}$ & $\mathrm{Len}_{22}$ \\ \midrule
$\tilde s^{(0,0)}$ (seed, model-first) & $\{f_8\}$                       & 1 & 1 (no edit) & 0.25 & 4.7k \\
$\tilde s^{(1,0)}$ (seed, test-first)  & $\{f_{11},f_5\}$                & 2 & 1 (no edit) & 0.58 & 8.7k \\
$s^{(1,f_3)}$ (curation on seed 1)     & $\{f_3,f_8,f_{9}\}$            & 3 & 1           & 0.70 & 5.4k \\
\rowcolor{highlightrow}
$s^{(0,f_3)}=s_{22}^\star$ (curation on seed 0) & $\{f_3,f_4,f_{10}\}$    & 3 & 1           & \textbf{0.70} & \textbf{4.3k} \\
\bottomrule
\end{tabular}
\end{table}

\noindent As a worked example of the $\mathrm{Eff}$ computation, take
$s^{(0,f_3)}$. The initial frontier is
$\mathcal{A}_{22}=\{f_3,f_5,f_8,f_{11}\}$ ($|\mathcal{A}|=4$). The
\emph{mutation step} itself realizes only $f_3$ ($\mathrm{Prog}=1/4=0.25$).
In the 7-step re-rolled suffix the blinded model -- now seeing that the
sibling handler uses \texttt{\_filters\_from\_querystring} -- naturally
opens \texttt{\_base\_response.py} to learn what the helper does, realizing
$f_4$ (frontier grew to $|\mathcal{A}|=5$ after $f_3$, so $1/5=0.20$); a
follow-up \texttt{grep -rn} for the pattern across
\texttt{moto/ec2/responses/} realizes $f_{10}$ ($1/4=0.25$). Summing the
three per-step ratios gives $\mathrm{Eff}_{22}\!\approx\!0.70$. The leakage
rule rejects $f_9$ in $\tilde s^{(0,0)}$: even though step~26's view of the
model file would expose it in the same observation as $f_8$, $f_9$'s
prerequisite ($f_8$) is not yet in $U_{\tau-1}$, so it contributes 0 and
can be realized only at a later window.

\paragraph{Single-curation construction.}
The curator scans positions $j\!=\!0,\ldots,9$ along seed $k\!=\!0$ and picks
$j\!=\!2$ (the first non-trivial branching point: the seed wastes step~25 on
an empty assistant thought). Among $\mathcal{A}_{24}=\mathcal{A}_{22}=\{f_3,f_5,f_8,f_{11}\}$
(steps $23$--$24$ of seed $k\!=\!0$ realize no new node; $f_8$ would be
realized only at step~26 of the unmutated seed via the model-file view, so
at $j\!=\!2$ it is still on the frontier), the $v\!=\!f_3$ has is chosen as the curation target node. The replacement response and its critic
verdicts are:

\begin{tcolorbox}[colback=blue!3, colframe=blue!40, sharp corners, boxsep=2pt,
                  left=4pt, right=4pt, top=2pt, bottom=2pt,
                  fontupper=\small, title=\textbf{Mutated step $y'_{22+2}$
                  (curator-authored, target $f_3$)}]
\textit{Assistant:} ``Before diving deeper into the model, let me first
check how the nearby \texttt{describe\_security\_groups} method in the same
response file parses its filters --- it's the closest sibling and should
reveal the standard pattern.''\\[2pt]
\textit{Tool call:}
\texttt{execute\_bash cmd="grep -n 'def describe\_security\_groups\textbackslash{}|filters' moto/ec2/responses/security\_groups.py | head -20"}
\end{tcolorbox}

\noindent\textbf{Symbolic referential-integrity check.} Entities mentioned
in $y'$ -- \texttt{describe\_security\_groups}, \texttt{filters},
\texttt{moto/ec2/responses/security\_groups.py}, the \texttt{grep} flag
\texttt{-n} -- all appear in $\mathrm{Obs}(h_{24})$ (the issue text and the
views at steps~22 and~24). \textbf{\ding{51}~PASS.}

\noindent\textbf{Neural claim-grounding judge.} The claim ``the sibling
method is in the same file and is a similar operation'' is entailed by the
view at step~22 (lines 190--210), where the tail of
\texttt{describe\_security\_groups} is visible just above the buggy method.
\textbf{\ding{51}~PASS.}\;\; Therefore $\mathrm{Ground}_{22}(s)=1$.
A counterfactual mutation targeting $f_5$ with the claim ``\texttt{\_get\_param}
does exact-key matching, which is why \texttt{Filter} fails'' would
\textbf{\ding{55}~FAIL} the neural judge -- that semantics has not yet been
observed -- and would zero out $\mathrm{Eff}$.

\begin{figure}[h]
\centering
\includegraphics[width=0.62\linewidth]{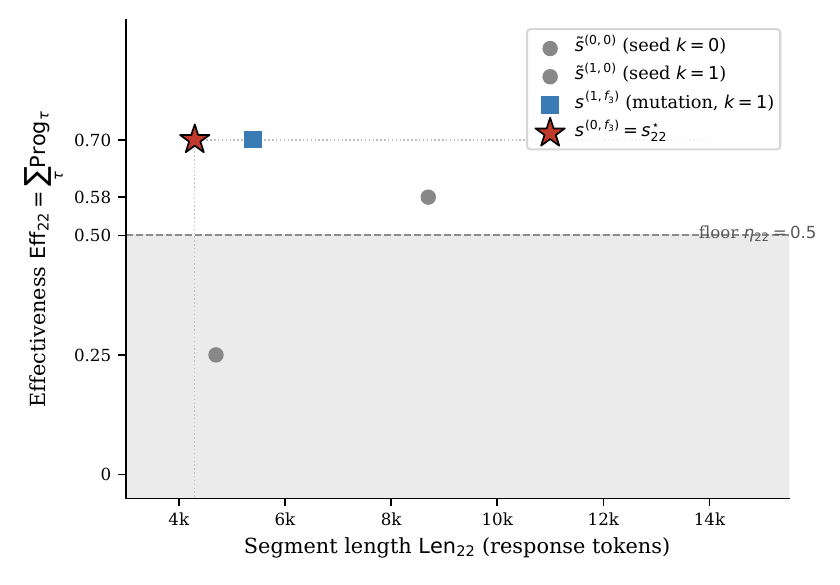}
\caption{Window-22 candidate pool in
$(\mathrm{Len}_{22},\mathrm{Eff}_{22})$ space, with $\mathrm{Len}_{22}$
measured in response tokens (assistant messages plus their observations).
Both pure seeds (grey circles) sit far below the floor
$\eta_{22}=0.50$ (shaded band, dashed line). The two mutated variants tie
at $\mathrm{Eff}=0.70$; the tie-break by length picks the seed-0 mutation
(red star), which is committed as $s_{22}^\star$.}
\label{fig:wex_pareto}
\end{figure}

\paragraph{Selection and commit.}
The two mutated variants both clear three new nodes and tie at
$\mathrm{Eff}_{22}=0.70$, well above the floor $\eta_{22}\!\approx\!0.5$,
but they realize \emph{different} parts of the graph:
$s^{(0,f_3)}$ extends along the response-handler chain
($\{f_3,f_4,f_{10}\}$, branching from $f_3$ into the helper definition and
its cross-file prevalence), whereas $s^{(1,f_3)}$ extends along the
model-layer chain ($\{f_3,f_8,f_9\}$, since the seed-1 prefix had already
oriented the agent toward the backend). Both pure seeds fall below the
floor. The two mutated variants thus form the local non-dominated set on
the $\mathrm{Eff}$ axis, and the tie-break by length selects the seed-0
mutation, which has the shorter prefix
($\mathrm{Len}=4.3$k vs $5.4$k tokens). $s^{(0,f_3)}$ is committed as
$s_{22}^\star$ (Fig.~\ref{fig:wex_pareto}). The first $\delta=4$ steps are
appended to the trajectory; the remaining suffix is replanned from the new
prefix. The non-selected candidates are rolled back and never enter the
trajectory.

\paragraph{Side-by-side comparison.}
Fig.~\ref{fig:wex_traj} contrasts the unmodified blinded rollout
(steps 23--32) with the \methodname trajectory
(steps 23--37) over the same window. The committed
prefix is the first $\delta=4$ steps of $s_{22}^\star$ -- step~23, step~24,
the mutated $y'$ at step~25, and the re-rolled step~26 that immediately
benefits from $y'$ -- and only the assistant message at step~25 is
curator-authored; every other committed token is a blinded continuation.
Follow-on windows then replan from the new prefix and converge on the
one-line response-handler fix, whereas the unmodified rollout commits to a
100-step backend over-engineering loop that edits \texttt{SecurityRule},
the response template, and breaks an existing test.

\begin{figure}[h]
\centering
\makebox[\linewidth][c]{%
\begin{tikzpicture}[
  font=\scriptsize\ttfamily,
  trajstep/.style={draw=gray!50, rounded corners=1pt, inner sep=1.5pt,
               minimum width=5.7cm, text width=5.5cm,
               minimum height=2.8mm, anchor=west,
               align=left, font=\scriptsize\ttfamily},
  asst/.style={trajstep, fill=gray!5},
  tool/.style={trajstep, fill=blue!4},
  mut/.style={trajstep, fill=red!12, draw=red!60, line width=0.6pt},
  noise/.style={trajstep, fill=red!5,  draw=red!30},
  good/.style={trajstep, fill=green!8, draw=green!50!black!40},
  hdr/.style={font=\small\bfseries\sffamily, anchor=west},
  arr/.style={->, >={Stealth[length=2pt]}, gray!60, line width=0.4pt},
]

% --- LEFT column (blinded) ---
\node[hdr] (lh) at (0, 0) {Blinded rollout (w/o curation)};
\node[asst,below=1.6mm of lh.south west,anchor=north west] (l23) {[23] A: Now let me check the model implementation};
\node[tool,below=0.6mm of l23.south west,anchor=north west] (l24) {[24] O: grep describe\_security\_group\_rules \(\to\) models:545};
\node[asst,below=0.6mm of l24.south west,anchor=north west] (l25) {[25] A: \emph{(empty thought)}};
\node[tool,below=0.6mm of l25.south west,anchor=north west] (l26) {[26] O: view models/security\_groups.py 540-560 \quad \(\to\) f8,f9};
\node[asst,below=0.6mm of l26.south west,anchor=north west] (l27) {[27] A: I see! filters=None passed to backend};
\node[noise,below=0.6mm of l27.south west,anchor=north west] (l28) {[28] O: re-view responses/security\_groups.py 195-205};
\node[asst,below=0.6mm of l28.south west,anchor=north west] (l29) {[29] A: Look at examples of filter parsing};
\node[noise,below=0.6mm of l29.south west,anchor=north west] (l30) {[30] O: grep Filter (30K chars of noise)};
\node[asst,below=0.6mm of l30.south west,anchor=north west] (l31) {[31] A: Trace SecurityRule for skipped filtering};
\node[noise,below=0.6mm of l31.south west,anchor=north west] (l32) {[32] O: view models/.../SecurityRule.\_\_init\_\_ \quad \(\to\) wrong path};
\node[font=\scriptsize\itshape, anchor=north west, text width=5.7cm,
      text=red!60!black, inner sep=0pt]
   at ([yshift=-1.5mm]l32.south west)
   {\(\downarrow\) commits to BACKEND fix; next 100 steps add group\_id to SecurityRule, edit response template, BREAK test\_create\_and\_describe\_security\_grp\_rule.};

% --- RIGHT column (curated) ---
\node[hdr,right=12mm of lh] (rh) {\methodname trajectory (with curation)};
\node[asst,below=1.6mm of rh.south west,anchor=north west] (r23) {[23] A: Examine responses/security\_groups.py};
\node[tool,below=0.6mm of r23.south west,anchor=north west] (r24) {[24] O: view responses/security\_groups.py 190-210 \(\to\) f2};
\node[mut,below=0.6mm of r24.south west,anchor=north west,
      minimum height=5.6mm, align=left]  (r25)
    {[25] A: \textcolor{gray!75}{\sout{Let me check the backend model implementation}} \;\textit{\textcolor{gray}{(blinded)}}\\%
     {[25] A: }Check sibling describe\_security\_groups parses $\;\leftarrow\;y'$ \textit{(curated)}};
\node[good,below=0.6mm of r25.south west,anchor=north west] (r26) {[26] O: grep \(\to\) line~186 \texttt{filters=\_filters\_from\_querystring()} \(\to\) \textbf{f3}};
\node[asst,below=0.6mm of r26.south west,anchor=north west] (r27) {[27] A: \_filters\_from\_querystring uses \_get\_multi\_param};
\node[good,below=0.6mm of r27.south west,anchor=north west] (r28) {[28] O: view \_base\_response.py 1-31 \(\to\) \textbf{f4}};
\node[asst,below=0.6mm of r28.south west,anchor=north west] (r29) {[29] A: Is this the standard EC2 pattern across responses?};
\node[good,below=0.6mm of r29.south west,anchor=north west] (r30) {[30] O: grep -rn \_filters\_from\_querystring responses/ \(\to\) \textbf{f10}};
\node[asst,below=0.6mm of r30.south west,anchor=north west] (r31) {[31] A: Plan: replace \_get\_param('Filter') with helper at line 197};
\node[tool,below=0.6mm of r31.south west,anchor=north west] (r32) {[32] O: \emph{(commits, replans from new prefix)}};
\node[font=\scriptsize\itshape, anchor=north west, text width=5.7cm,
      text=green!40!black, inner sep=0pt]
   at ([yshift=-1.5mm]r32.south west)
   {\(\downarrow\) one-line edit at responses/security\_groups.py:197; both validations pass.};

\end{tikzpicture}%
}
\caption{Side-by-side fragment of the same window in $\pi_{\mathrm{blind}}$
(left) and \methodname (right). Rows labelled \texttt{A} are assistant
responses; rows labelled \texttt{O} are environment observations. The single
curator-authored step (red,
$y'$ at $j{=}2$) replaces an empty assistant thought with a frontier-aware
inspection of the sibling \texttt{describe\_security\_groups} method,
unlocking $f_3$. The 7-step re-rolled blinded suffix then organically
realizes only the two facts naturally implied by that observation
($f_4,f_{10}$, green steps), whereas the unmodified rollout (red steps)
wanders into the model layer and breaks an existing test.}
\label{fig:wex_traj}
\end{figure}

\section{Training details}
\label{app:training_details}

\paragraph{Student SFT.}
We fine-tune Qwen2.5-Coder-14B/32B-Instruct~\citep{qwen25coder} with \textbf{ms-swift}'s Megatron-LM backend~\citep{ms-swift}. Full-parameter SFT runs for $3$ epochs with Adam (lr $2\!\times\!10^{-5}$, $\beta_1=0.9$, $\beta_2=0.999$, weight decay $0.01$, gradient clipping $1.0$), global batch size $4$ with sequence packing up to $131{,}072$ tokens, linear warm-up over the first $5\%$ of steps and cosine decay to $0$, and BF16 mixed precision. The loss is computed on assistant tokens only.

\paragraph{Context window and parallelism.}
Curated trajectories frequently exceed the base $32{,}768$-token window of Qwen2.5-Coder; we extend the effective context to $131{,}072$ tokens via YaRN positional interpolation~\citep{peng2023yarn} (factor $4$) for both training and inference. Training uses tensor parallelism $2$, context parallelism $4$, and Megatron-style sequence parallelism.

\paragraph{Compute resources.}
All experiments---curation, SFT training, and SWE-bench evaluation---are run on a single node with $8\times$ NVIDIA H200 (141\,GB) GPUs. Teacher rollouts and student inference are served via vLLM on the same node, and the curation pipeline (reverse-phase graph distillation and forward-phase trajectory construction) is executed on the same hardware.

\section{Compute-matched rejection-sampling baseline}
\label{app:compute_matched}

The size-matched control in Sec.~\ref{sec:rq1} fixes the number of curated trajectories but ignores the GPU-hours \methodname spends on graph distillation, blinded seeding, mutation, and gating. A skeptical reader may ask whether the same compute, redirected into additional plain rollouts, would close the gap. This appendix runs that experiment under a strict compute-parity protocol on the same hardware (single $8\times$H200 node, vLLM-served Qwen3-Coder-480B teacher, OpenHands scaffold, $100$-iteration ReAct budget).

\paragraph{Compute accounting.}
End-to-end \methodname curation over the $1.8$k SWE-Gym instances costs $\approx\!226$ GPU-hours (reverse-phase proposer/critic + forward-phase blinded seeds, mutations, rollbacks, and per-step LLM gating). One full pass of plain teacher rollouts over the same SWE-Gym pool costs $\approx\!68$ GPU-hours. We therefore allocate the rejection-sampling baseline a budget of $4$ rollout passes ($\approx\!272$ GPU-hours, $\sim\!20\%$ \emph{more} than \methodname), aggregate the resolved trajectories across passes (deduplicating per instance by keeping the shortest passing trajectory), and SFT Qwen2.5-Coder-32B-Instruct on the result under the same recipe as Sec.~\ref{sec:exp_setup}. As an upper bound on the supervision the $4$-run baseline could in principle harvest, the cumulative resolve rate (Pass@4) over the $4$ rollout passes on the SWE-Gym training pool reaches $34.3\%$, vs.\ a per-run Pass@1 of $\sim\!29\%$; the additional passes thus recover only a sub-set of new instances rather than uniformly improving every trajectory.

\paragraph{Results.}
Table~\ref{tab:compute_matched} compares the compute-matched RS baseline against \methodname (full) on SWE-bench Verified, reporting Pass@1, per-instance inference cost, and total curation GPU-hours.

\begin{table}[h]
    \centering
    \caption{\textbf{Compute-matched comparison on SWE-bench Verified} (Qwen3-C-480B teacher, Qwen2.5-C-32B student). The RS baseline is allocated $4$ rollout passes ($\approx\!272$ GPU-hours, $\sim\!20\%$ more than \methodname's curation budget); resolved trajectories are aggregated across passes (shortest passing kept per instance) and used as SFT data under the same recipe as Sec.~\ref{sec:exp_setup}. \textbf{Bold} denotes the better cell.}
    \label{tab:compute_matched}
    \renewcommand{\arraystretch}{1.2}
    \begin{tabular}{@{} l c c c @{}}
        \toprule
        \textbf{Recipe} & \textbf{Pass@1 (\%) $\uparrow$} & \textbf{Cost (\$) $\downarrow$} & \textbf{Curation GPU-hours $\downarrow$} \\
        \midrule
        Test-Pass RS ($4\!\times$ rollouts) & 43.2 & 0.86 & $\approx 272$ \\
        \rowcolor{gray!15}
        \methodname (full)                  & \textbf{50.4} & \textbf{0.78} & $\approx \mathbf{226}$ \\
        \bottomrule
    \end{tabular}
\end{table}

Even with $\sim\!20\%$ more curation compute, the rejection-sampling baseline remains bounded by its own ceiling: the additional rollout passes recover supervision only on instances the teacher happens to solve at least once across $4$ tries, and contribute nothing to the per-step quality of the trajectories that do get retained. \methodname spends comparable GPU-hours but allocates them differently---toward distilling $G^\star$ and shaping each retained trajectory along $(\mathrm{Eff},\mathrm{Len})$---which is what unlocks the simultaneous Pass@1 gain and per-instance inference-cost reduction in Table~\ref{tab:main_results}.

\section{Limitations and outlook}
\label{app:limitations}

\paragraph{Limitations.}
\methodname inherits four constraints worth flagging. (i)~\emph{Privileged-signal availability.} The reverse phase requires a reference patch $p^\star$ and an executable test suite per instance. Both hold on SWE-Gym and SWE-bench but exclude issues without a maintained CI; extending the recipe to such issues would require a surrogate verifier (e.g.,\ teacher-generated patches gated by self-consistency or property-based tests). (ii)~\emph{LLM-mediated guarantees.} Both the non-leakage critic in Phase~1 and the groundedness/establishment judges in Phase~2 are LLM-instantiated. Our value-of-information and component ablations (App.~\ref{app:voi}, Sec.~\ref{sec:rq4}) show the resulting graphs and gates carry the right signal in aggregate, but per-instance correctness is empirical rather than formal; a tighter, model-agnostic certificate of minimality and grounding remains open. (iii)~\emph{Curation compute.} Distillation and forward realization cost on the order of $10^2$ GPU-hours for $1.8$k instances (App.~\ref{app:compute_matched}); while compute-matched against rollout scaling, this is non-trivial and scales with graph size. (iv)~\emph{Scaffold and student scope.} Experiments fix OpenHands as the scaffold and the Qwen2.5-Coder family as the student; transfer to agentless pipelines, browser-augmented agents, or non-Qwen students is not yet validated.

\paragraph{Outlook.}
Two extensions follow naturally. First, $G^\star$ is a generic latent representation of process supervision: training a process reward model to score per-step compliance with $G^\star$ would lift the signal from behavior cloning to online RL while preserving non-leakage, addressing a known weakness of SFT under distribution shift. Second, the privileged-signal $\to$ latent-structure $\to$ blinded-realization factorization is not specific to code; any task with a verifiable terminal artifact and a partially observable, tool-mediated solution process (theorem proving, scientific protocol execution, multi-step data analysis) admits the same recipe. Together, these directions suggest that the right unit of supervision for capable agents is neither the trajectory nor the terminal artifact in isolation, but the \emph{structure} that links them.

%%%%%%%%%%%%%%%%%%%%%%%%%%%%%%%%%%%%%%%%%%%%%%%%%%%%%%%%%%%%

\newpage